\documentclass[usenatbib]{mn2e}

\usepackage{natbibmnfix}
\usepackage{astrojournals}
\usepackage{graphicx}
\usepackage{footmisc}
\usepackage{amssymb}
\usepackage{amsmath} 
\usepackage[varg]{txfonts}

\newcommand{\xmm}{\textit{XMM-Newton }}
\newcommand{\gmm}{$\Gamma$ }  
\newcommand{\nh}{$N\sb{\rm H}$ }  
\newcommand{\av}{$A\sb{\rm V}$ }  
\newcommand{\lumx}{$L\sb{\rm X}$ }
\newcommand{\loglumx}{${\rm log}\,(L\sb{\rm X})$ } 
\newcommand{\lbol}{$L\sb{\rm bol}$ } 
\newcommand{\loglbol}{${\rm log}\,(L\sb{\rm bol})$ }
\newcommand{\eddratio}{$\lambda\sb{\rm Edd}$ }
\newcommand{\mbh}{$M\sb{\rm BH}$ } 
\newcommand{\logmbh}{${\rm log}\,(M\sb{\rm BH})$ }

\newcommand{\wise}{\textit{WISE }}
\newcommand{\galex}{\textit{GALEX }}
\newcommand{\aox}{$\alpha\sb{\rm OX}$ }
\newcommand{\rl}{$R\sb{\rm L}$ }
\newcommand{\angstrom}{\text{\normalfont\AA}}
\newcommand{\mdot}{\dot m}
\newcommand{\msol}{M_\odot}
\newcommand{\lya}{Ly$\alpha$ }

\title[SEDs of type 1 AGN]{Do the spectral energy distributions of type 1 AGN show diversity?}
\author[A.\ E.\ Scott \& G.\ C.\ Stewart]{A.\ E.\ Scott\thanks{E-mail: amyscott@psu.edu} and G.\ C.\ Stewart\\
Department of Physics and Astronomy, University of Leicester, University Road, Leicester LE1 7RH, UK}

\begin{document}

\date{Accepted 2013 December 03. Received 2013 November 29; in original form 2013 May 01}

\pagerange{\pageref{firstpage}--\pageref{lastpage}} \pubyear{2013}

\maketitle

\label{firstpage}


\begin{abstract}
We create broadband spectral energy distributions (SEDs) of 761 type 1
active galactic nuclei (AGN).  The Scott et al. sample, created by a
cross-correlation of the optical SDSS DR5 quasar catalogue and the
2XMMi catalogue of serendipitous X-ray sources, is further matched
with the FIRST catalogue of radio sources, the mid-infrared (MIR)
\wise all-sky data release, the 2MASS near-infrared point source
catalogue, the UKIDSS DR9 Large Area Survey and the \galex all-sky and
medium ultraviolet imaging surveys.  This allows broadband SEDs
including up to 19 flux measurements covering $\textrm{log}\,\nu
\sim\,9.2-18.1$ to be created.  We investigate variations in the SED
shape by binning a subsample of 237 AGN with the best quality SEDs
according to their X-ray spectral parameters, their quasar sub-type
and physical parameters such as luminosity, black hole mass and
Eddington ratio.  The AGN sub-populations show some significant
differences in their SEDs; X-ray absorbed AGN show a deficit of
emission at X-ray/UV frequencies and an excess in the MIR consistent
with absorption and re-emission, radio-loud AGN show increased radio
and X-ray emission, consistent with the presence of a jet component in
addition to the emission seen from radio-quiet AGN and the SEDs of
Narrow-line Seyfert 1s only differ from other type 1s in the X-ray
regime, suggesting any physical differences are limited to their X-ray
emitting region.  Binning the AGN according to underlying physical
parameters reveals more subtle differences in the SEDs.  The X-ray
spectral slope does not appear to have any influence or dependence on
the multiwavelength emission in the rest of the SED.  The contribution
of X-rays to the bolometric luminosity is lower in higher luminosity
sources, and relatively more emission in the optical/UV is seen in AGN
with higher X-ray luminosities.  Variations in the relative flux and
peak frequency of the big blue bump are observed and may suggest
higher inner disc temperatures with increasing accretion rates.
Overall, we find that the diversity in the SED shapes is relatively
small, and we find no apparent single driver for the variations.
\end{abstract}




\section{Introduction}
\label{section:intro}
Active Galactic Nuclei (AGN) have high bolometric luminosities
($10\sp{44}-10\sp{48}\,\textrm{erg s}\sp{-1}$) and emit across the
entire electromagnetic spectrum.  The amount of energy output at
different frequencies is described by a spectral energy distribution
(SED).  It shows features originating from separate physical processes
occurring in different regions around the central, supermassive black
hole.  The radio emission of AGN contributes very little to the
bolometric output, even in radio-loud quasars (RLQ) where it is
typically $100-1000$ times greater than that of radio-quiet quasars
(RQQ) and originates from synchrotron emission in their relativistic
jets.  A broad `hump' of IR emission at $2-25\,\mu\textrm{m}$ is
thought to be due to thermal re-processing by dust surrounding the
black hole \citep{rees69,rieke78}, possibly in the torus
\citep{antonucci85}.  The UV and optical emission from AGN is
dominated by the `big blue bump' (BBB), which is attributed primarily
to thermal emission from the accretion disc \citep{shields78,
  malkan82, czerny87}.  The X-ray emission is thought to be produced
mainly by the inverse Compton scattering of low energy UV photons from
the accretion disc by relativistic electrons in the corona
\citep{haardt93}.  It appears in the spectrum as a simple power law
described by $P\sb{\rm E}\propto E\sp{-\Gamma}$ where \gmm is the
photon index \citep{mushotzky80} and may be modified by reflection in
the accretion disc \citep{pounds90}.  $\gamma$-rays have also now been
detected from a number of AGN \citep{wagner08}, but contribute very
little to the total luminous output in non-blazar type 1 AGN.  An
accurate knowledge of the relative strengths and shapes of the
different components in AGN SEDs can aid the understanding of the
physical processes occurring in the object.  In particular,
understanding how the SED shape changes with underlying physical
parameters such as the mass of the black hole, $M\sb{\rm BH}$, and the
accretion rate, can help us to better understand the overall accretion
process.

Constructing complete SEDs requires data in multiple wavebands,
necessitating observations from many different instruments and
observatories, both space-borne and ground based.  Consequently this
process has historically been very time consuming.  The seminal work
which created full SEDs for 47 quasars (29 RQQ and 18 RLQ) was that of
\citet{elvis94} and included X-ray to radio data.  Variation in the
SEDs of different objects was found, and the mean energy distribution
was presented.  With the greater availability of multiwavelength data,
larger samples of SEDs can now be created.  \citet{richards06}
constructed SEDs of 259 quasars using photometry from the X-ray
through to the radio band.  More recently, \citet{lusso12} have
constructed SEDs of 545 X-ray selected type 1 AGN, and \citet{elvis12}
present a mean SED of 413 type 1 AGN, both from the
\textit{XMM}-COSMOS deep field.  \citet{shang11} have created SEDs of
a smaller sample of objects, with a complicated selection function,
but they include high resolution UV/optical and MIR spectra in
addition to photometry in the radio, FIR and NIR bands.

\citet{jin12a} studies the optical to X-ray SEDs of a small sample of
51 type 1 AGN, including 12 Narrow-line Seyfert 1 (NLS1), using high
quality spectra from the Sloan Digital Sky Survey (SDSS;
\citealt{SDSS}) and \textit{XMM-Newton}.  They fit their partial SEDs
with a physically motivated model incorporating the disc emission,
comptonisation and the X-ray power-law, and derive physical quantities
such as the power-law slope, $\Gamma$, the $2-10\,\textrm{keV}$ X-ray
luminosity, $L\sb{\rm X}$, the bolometric luminosity, $L\sb{\rm bol}$,
bolometric correction factors, $\kappa$, the Eddington ratio,
$\lambda\sb{\rm Edd}$, and $\alpha\sb{\rm OX}$\footnote{$\alpha\sb{\rm
    OX}=0.38\,\textrm{log}\left(f\sb{\rm X}/f\sb{\rm O}\right)$ where
  $f\sb{\rm X}$ is the X-ray flux at $2\,\textrm{keV}$ and $f\sb{\rm
    O}$ is the optical flux at $2500\,\textrm{\AA}$.\label{aox}},
directly from the model fit.  Mean SEDs were presented, binned
according to their physical parameters.  It was found that the SED
changes are similar for most parameters except $L\sb{\rm bol}$, likely
because the SED shape does not depend solely on one key parameter but
ultimately depends on a combination of \eddratio and $M\sb{\rm BH}$.
\citet{hao12} investigate how the NIR--UV SED shape varies with
respect to $z$, $L\sb{\rm bol}$, $M\sb{\rm BH}$ and $\lambda\sb{\rm
  Edd}$ using a sub-sample of 200 radio-quiet type 1 AGN from the
\textit{XMM}-COSMOS survey.  They find no dependence of the SED shape
on these parameters.  \citet{krawczyk13} create a mean MIR--UV SED of
$108\,184$ type 1 AGN and investigate the SED dependence on UV
parameters.  They find that the SEDs for lower UV luminosity AGN show
harder $\alpha\sb{\rm UV}$, redder optical continuua and lower amounts
of hot dust.

In this work we create broadband SEDs for the sample of 761 type 1 AGN
described in \citeauthor{scott11} (\citeyear{scott11}; hereafter S11).
The sample was created by a cross-correlation of the serendipitous
X-ray source catalogue 2XMMi \citep{2XMM} and the optical SDSS Data
Release 5 quasar catalogue \citep{DR5QSO}.  We also include photometry
in the radio, MIR, NIR and UV bands.  The sample includes sources
representative of typical type 1 AGN, and the large number allows us
to investigate how the SED shape changes with physical parameters
spanning a large range of values.  We construct average broadband
SEDs, binned according to their X-ray spectral properties, \gmm and
the inclusion of spectral components, redshift, their quasar sub-type
including $R\sb{\rm L}$\footnote{$R\sb{\rm L}=f\sb{\rm R}/f\sb{\rm O}$
  where $f\sb{\rm R}$ is the radio flux at $5\,\textrm{GHz}$ and
  $f\sb{\rm O}$ is the optical flux at
  $4400\,\textrm{\AA}$.\label{rl}}, and the physical parameters
$L\sb{\rm X}$, $L\sb{\rm bol}$, $M\sb{\rm BH}$ and $\lambda\sb{\rm
  Edd}$.


\section{Data}
\label{section:data}
\begin{figure}
  \centering
  \includegraphics[width=0.48\textwidth]{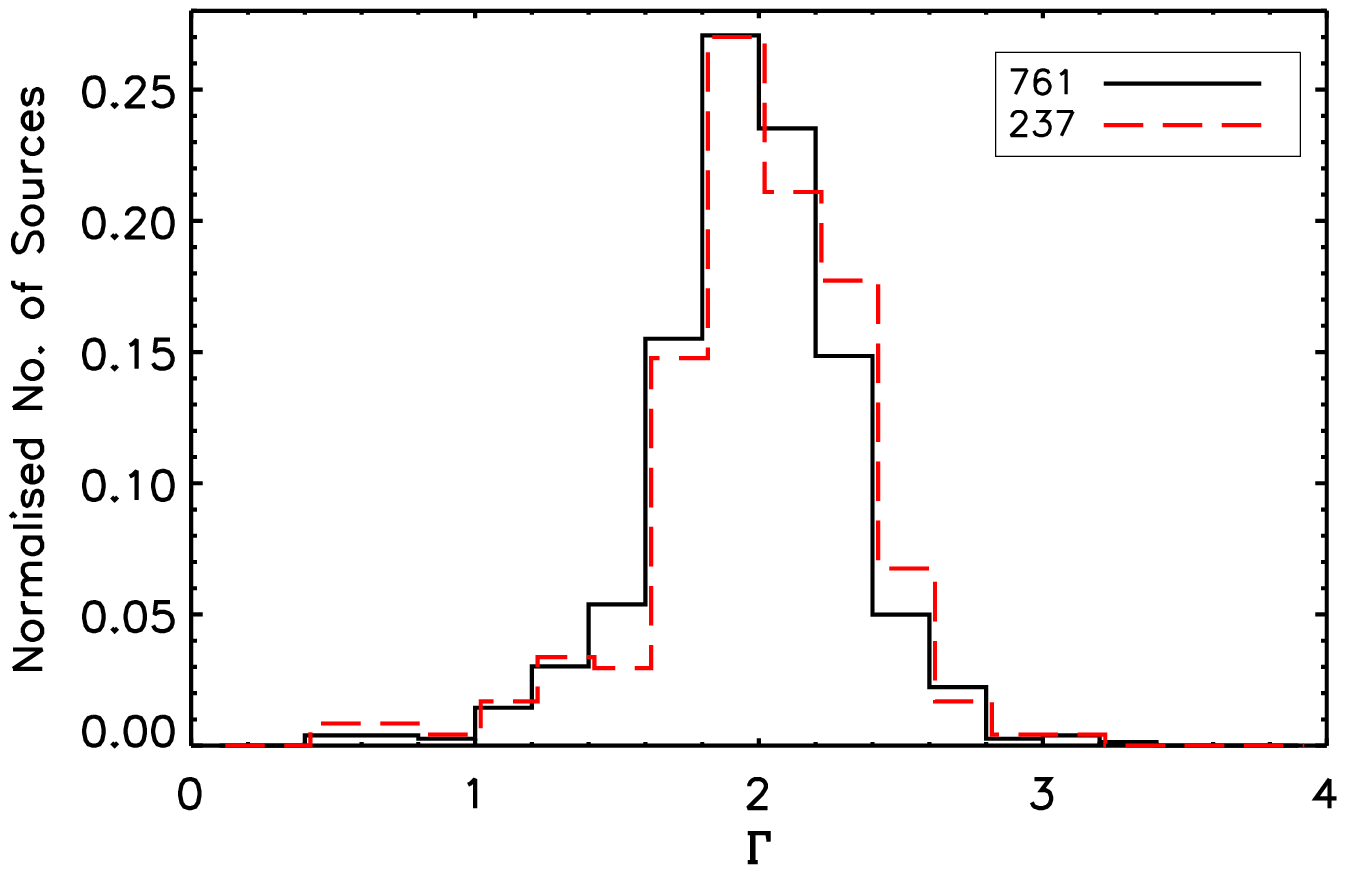}\\
  \includegraphics[width=0.48\textwidth]{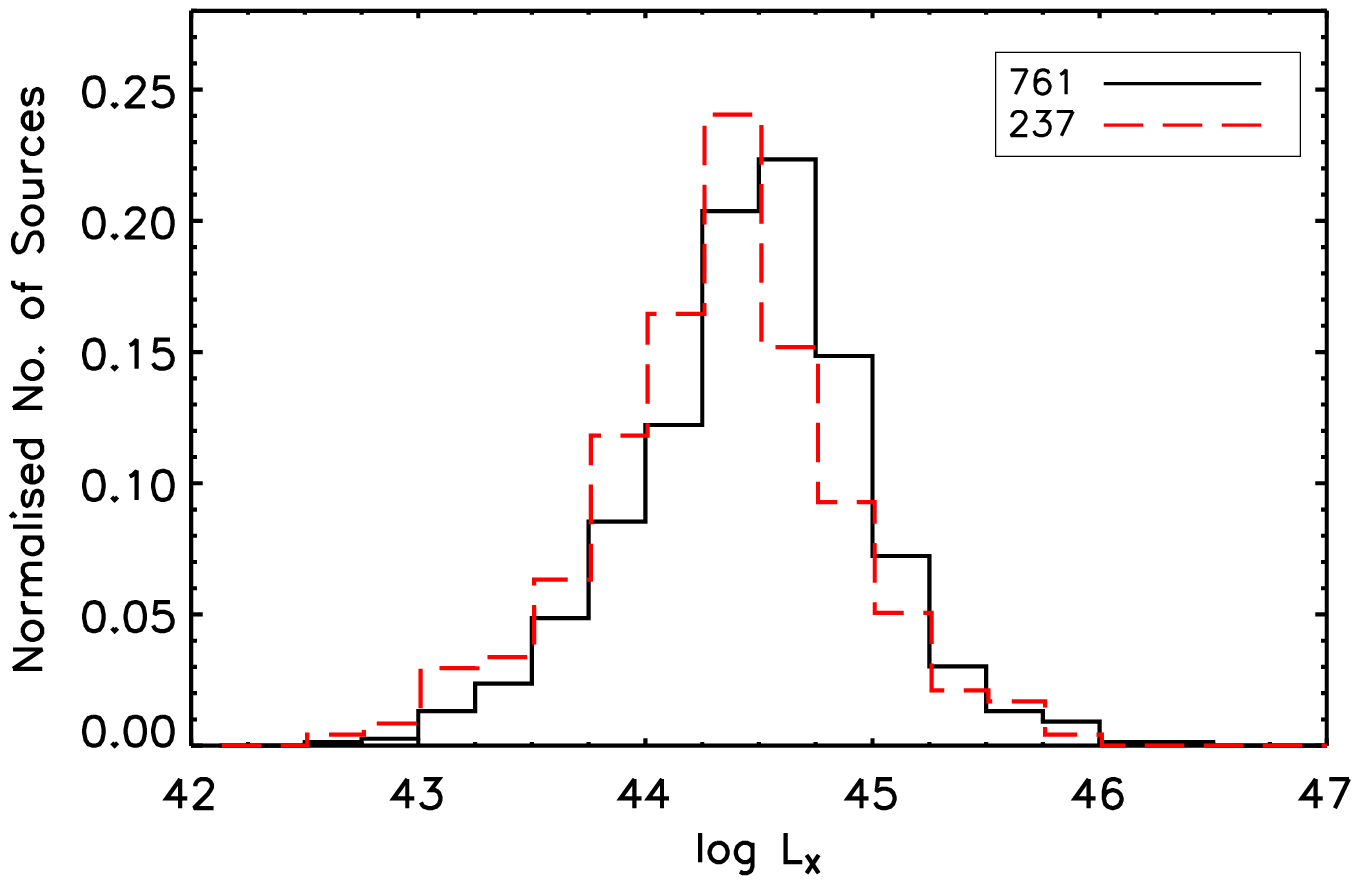}
  \caption{Shown with the black, solid lines are the distributions of
    the best-fitting power-law slope, \gmm (top) and 2--10 keV X-ray
    luminosity (bottom), which were first presented in S11.  These are
    compared to the distributions for the subsample of 237 sources
    considered later in this paper (red, dashed), plotted with a
    small, arbitrary offset for clarity.}
  \label{fig:s11}
\end{figure}

This paper studies the multiwavelength properties of a large sample of
761 type 1 AGN, for which a comprehensive X-ray spectral analysis was
presented in S11.  This included estimates of the power-law slope \gmm
from the best-fitting spectral model, and estimates of the unabsorbed
2-10 keV X-ray luminosity, $L\sb{\rm X}$.  The distributions of these
2 parameters are shown in black in Fig.~\ref{fig:s11}.  For the
analysis in this paper we estimate the X-ray flux in narrow bands
using the best-fitting spectral model, and convert them to
monochromatic flux estimates at $0.5$, $1.0$ and $5.0$ keV.  These
values are corrected for Galactic absorption, assuming an \nh value
taken from the H\textsc{i} map of \citet{dickey90}.

Each of the sources has optical photometry in the $5$ SDSS
\textit{ugriz} bands.  The catalogue PSF magnitudes are first
corrected for Galactic extinction using values from the maps of
\citet{schlegel98} and then converted into the AB system
\citep{okegunn83}.  The magnitudes (and their associated errors) are
converted into a flux density by Equation~\ref{equation:flux_mag}
where $F\sb{0}=3631$ Jy.

\begin{equation}
\textrm{Flux (Jy)} = F\sb{\rm 0} \times 10\sp{-0.4m}
\label{equation:flux_mag}
\end{equation}

Radio fluxes for the sources at 1.4 GHz were obtained from the FIRST
catalogue \citep{FIRST}.  A $20''$ matching radius was used in order
to ensure that any extended emission from the source was included, and
the radio fluxes from any multiple matches were summed. 104 sources
had a detection in FIRST and for the 613 sources with no detection,
$5\,\times$ the RMS at the source position was used as an upper
limit. 44 sources did not lie in the sky area covered by the FIRST
survey.  For sources with a detection, the error on the integrated
flux was taken to be the RMS value at the source position.  In the
case of multiple detections, the RMS values for each are added in
quadrature.

Near-Infrared (NIR) fluxes were obtained from the 2MASS Point Source
Catalogue \citep{2MASSPSC} using a $2''$ matching radius.  $166$
matches were found, with no sources having duplicate detections.  Only
two sources have separations greater than $1$\arcsec, and both are
less than $1.5$\arcsec.  Approximately $75$ per cent of the source
detections are part of the `faint extension' and have magnitudes as
faint as $17.6$\,(J), $17.8$\,(H) and $16.3$\,(K).  The magnitudes and
corresponding errors given in the catalogue were converted to fluxes
using Equation~\ref{equation:flux_mag} where $F\sb{\rm 0}$ for each of
the 3 bands is; J: $F\sb{\rm 0}=1594\pm27.8{\rm \,Jy}$, H: $F\sb{\rm
  0}=1024\pm20.0{\rm \,Jy}$, K: $F\sb{\rm 0}=666.7\pm 12.6{\rm \,Jy}$.
The NIR data were excluded for sources with catalogue flags indicating
poor photometry.  The sample was also cross-correlated with the UKIDSS
survey \citep{lawrence07}, which uses the Wide-field camera
(\citealt{casali07}; WFCAM) on UKIRT.  The Large Area Survey (LAS)
from DR9 was searched for matches within a $1''$ radius of the optical
source positions.  353 sources had data in at least one of the 4
UKIDSS photometric bands YJHK \citep{hewett06,hodgkin09}.  The default
point-source aperture-corrected magnitudes and errors were extracted
from the archive \citep{hambly08} and were converted to fluxes using
Equation~\ref{equation:flux_mag} where $F\sb{\rm 0}$ for each of the 4
bands is; Y: $F\sb{\rm 0}=2026{\rm \,Jy}$, J: $F\sb{\rm 0}=1530{\rm
  \,Jy}$, H: $F\sb{\rm 0}=1019{\rm \,Jy}$, K: $F\sb{\rm 0}=631{\rm
  \,Jy}$.  For 72 sources NIR data is available from both 2MASS and
the deeper UKIDSS survey.  In these cases we use the UKIDSS data in
preference to any 2MASS data for individual bands, due to the errors
on the photometry being an order of magnitude lower.  The dispersion
between the flux estimates from each survey is $\lesssim 0.2$ mag.
Combining both data sets, 408 sources have a flux estimate in the JHK
bands (309 also include a Y band flux).

The sample was cross-correlated with the All Sky Data Release
\citep{wise_all_sky} from \textit{WISE} to obtain mid-infrared (MIR)
fluxes.  A $5''$ matching radius was used giving 819 matches
corresponding to 754 unique objects.  In cases where two \wise
detections were matched to the same SDSS coordinates, the one with the
largest separation was excluded, provided the closest match had a
separation $<1''$.  Any further double matches within the $6''$
resolution limit were excluded.  This gave a final sample where 754
sources have a \wise detection, $98$ per cent of which have a
separation $<2''$ from the optical position.  The four broad band
\wise magnitudes and associated errors were converted to fluxes by
Equation~\ref{equation:flux_mag} using the zero point fluxes W1:
$F\sb{\rm 0} = 306.682 \pm 4.600{\rm \,Jy}$, W2: $F\sb{\rm 0} =
170.663 \pm 2.600{\rm \,Jy}$, W3: $F\sb{\rm 0} = 29.045 \pm 0.436{\rm
  \,Jy}$ and W4: $F\sb{\rm 0} = 8.284 \pm 0.124{\rm \,Jy}$.  The \wise
data was excluded if the photometry was flagged as poor by the
catalogue.  46 sources were affected in at least one of the \wise
bands, but only 16 unique sources were affected in all four.

Ultraviolet (UV) fluxes were obtained from \galex GR6 data using
CASJobs\footnote{http://galex.stsci.edu/casjobs/}.  A matching radius
of $2''$, as recommended by \citet{galex}, yielded 1664 matches to 645
unique sources.  We limit the detection list to those from AIS
(All-sky Imaging Survey) or MIS (Medium Imaging Survey) observations
as their magnitude limits are well defined.  We use the MIS detection
where both are available.  We calculate the likelihood ratio, $L$, and
the reliability, $R$, of each detection being the correct match
(\citealt{sutherland92}; Equation~\ref{equation:likelihood}) and
choose the detection with the highest value of $R$ as the most
appropriate.  All detections have $R>50$ per cent.

\begin{equation}
  L = \frac{Q(<m) \textrm{exp}(-r\sp{2}/2)}{2 \pi \sigma\sb{1} \sigma\sb{2} N(<m)};
  R\sb{j} = \frac{L\sb{j}}{\sum{L\sb{i}} + [1 - Q(<m)]}
\label{equation:likelihood}
\end{equation}

This gives a final sample of 590 unique sources with a UV detection.
The UV magnitudes are corrected for Galactic reddening using the
reddening law $A\sb{UV} = R\sb{UV}\,E(B-V)$ \citep{cardelli89} where
the $E(B-V)$ values were obtained from the maps of \citet{schlegel98}
and $R\sb{\rm UV}=8.24$ \citep{wyder07}.  UV fluxes are calculated
using Equation~\ref{equation:flux_mag} where $F\sb{\rm 0}=3631\,{\rm
  Jy}$.  358 sources have a detection in both the NUV and FUV bands,
224 sources only have a detection in the NUV and 8 have FUV data only.

Table~\ref{table:coverage} lists the percentage of the sample which
have a flux measurement in each of the wavebands considered in this
analysis.

\begin{table}
\centering
  \caption{The observed frequencies of the $19$ wavebands used in the
    multiwavelength catalogues and the percentage of the $761$ sources
    with a flux measurement in each.  Upper limits which are not used
    in the SED creation and detections with bad photometry are not
    included in these numbers.  \textit{Notes:} $^a$These wavelengths
    correspond to the UKIDSS survey; $^b$ correspond to the 2MASS
    survey.}
  \label{table:coverage}
     \begin{centering}
     \begin{tabular}{llccc}
     \hline
     \multicolumn{3}{c}{Waveband} &
     \multicolumn{1}{c}{log $\nu$} &
     \multicolumn{1}{c}{Percentage} \\[0.5ex]
     \hline
     X-ray                            & Hard       & $5\,{\rm keV}$         & $18.08$  & $100$    \\[0.5ex]
     \hspace*{0.2cm}(\textit{XMM})    & Soft       & $1\,{\rm keV}$         & $17.38$  & $100$    \\[0.5ex]
                                      & Very soft  & $0.5\,{\rm keV}$       & $17.08$  & $ 98$    \\[1.0ex]

     UV                               & FUV        & $1539{\rm \AA}$        & $15.29$  & $48$     \\[0.5ex]
     \hspace*{0.2cm}(\textit{GALEX})  & NUV        & $2316{\rm \AA}$        & $15.11$  & $76$     \\[1.0ex]

     Optical                          & $u$        & $3500{\rm \AA}$        & $14.93$  & $100$    \\[0.5ex]
     \hspace*{0.2cm}(SDSS)            & $g$        & $4800{\rm \AA}$        & $14.80$  & $100$    \\[0.5ex]
                                      & $r$        & $6250{\rm \AA}$        & $14.68$  & $100$    \\[0.5ex]
                                      & $i$        & $7700{\rm \AA}$        & $14.59$  & $100$    \\[0.5ex]
                                      & $z$        & $9100{\rm \AA}$        & $14.52$  & $100$    \\[1.0ex]

     NIR                              & Y          & $1.03\mu{\rm m}^a$     & $14.46$  & $47$     \\[0.5ex]
     \hspace*{0.2cm}(UKIDSS$^a$       & J          & $1.25\mu{\rm m}^{a,b}$ & $14.38$  & $58$     \\[0.5ex]
     \hspace*{0.25cm}\& 2MASS$^b$)    & H          & $1.63\mu{\rm m}^a$     & $14.26$  & $59$     \\[0.5ex]
                                      &            & $1.65\mu{\rm m}^b$     & $14.26$  &          \\[0.5ex]
                                      & K          & $2.20\mu{\rm m}^a$     & $14.13$  & $60$     \\[0.5ex]
                                      &            & $2.17\mu{\rm m}^b$     & $14.14$  &          \\[1.0ex]

     MIR                              & W1         & $3.4\mu{\rm m}$        & $13.95$  & $94$     \\[0.5ex]
     \hspace*{0.2cm}(\textit{WISE})   & W2         & $4.6\mu{\rm m}$        & $13.81$  & $95$     \\[0.5ex]
                                      & W3         & $12 \mu{\rm m}$        & $13.40$  & $90$     \\[0.5ex]
                                      & W4         & $22 \mu{\rm m}$        & $13.13$  & $65$     \\[1.0ex]

     Radio                            &            & $1.4\,{\rm GHz}$       &  $9.16$  & $14$     \\[0.5ex]
     \hspace*{0.2cm}(FIRST)           &            &                        &          &          \\[0.5ex] 
     \hline
     \end{tabular}
     \end{centering}
\end{table}


\section{Average Spectral Energy Distributions}
\label{section:seds}
\begin{figure}
  \centering
  \includegraphics[width=0.48\textwidth]{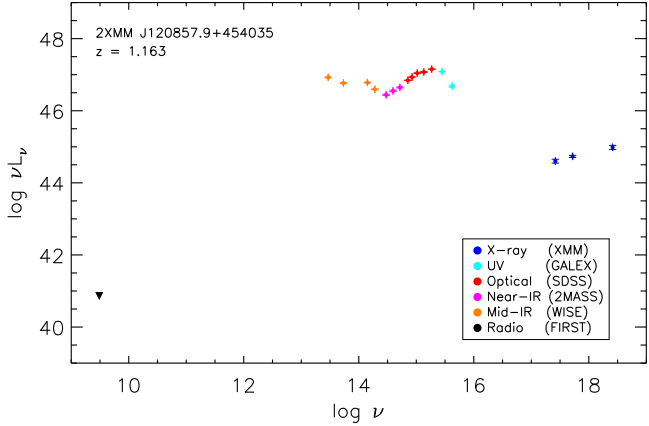}
  \caption{The rest-frame spectral energy distribution of 2XMM
    J120857.9+454035.  It includes radio data from FIRST (black), MIR
    data from \wise (orange), NIR data from 2MASS (pink), optical data
    from SDSS (red), UV data from \galex (light blue) and X-ray data
    from \xmm (dark blue).  Filled circles with error bars correspond
    to flux measurements.  Triangles denote upper limit estimates.}
  \label{fig:example_sed}
\end{figure}

Rest-frame monochromatic luminosities are calculated at each
wavelength and for each individual object using $\nu L\sb{\rm \nu} = 4
\pi d\sb{\rm L}\sp{2} \nu'F'\sb{\nu}$ where $\nu'$ is the observed
frequency, $F'\sb{\nu}$ is the observed flux and $d\sb{\rm L}$ is the
luminosity distance calculated from the redshift of the source
assuming a flat cosmology with ${\rm H}\sb{0}=70\,{\rm
  km\,s}\sp{-1}\,{\rm Mpc}\sp{-1}$, $\Omega\sb{\rm M}=0.3$ and
$\Omega\sb{\Lambda}=0.7$ \citep{spergel03}.  An example of a single
AGN SED plotted using ${\rm log}\,\nu$ on the $x$ axis (in units of
log Hz) and ${\rm log}\,\nu L\sb{\rm \nu}$ on the $y$ (in units of log
erg s$\sp{-1}$) is shown in Fig.~\ref{fig:example_sed}.  Luminosity
measurements with errors are plotted in different colours depending on
the source of the data.  Measurements with bad photometry flags are
excluded completely, whilst upper limits are plotted as triangles.  It
shows many of the typical features expected in the SED of a type 1 AGN
including: Power-law emission in the X-ray regime, in this case with a
flat spectral slope of $\Gamma\!\sim\!1.6$, a gap in the SED between
the X-ray and UV emission which cannot be sampled due to absorption by
the Milky Way, the `big blue bump' at (log) frequencies of
$14.5-15.5$, an inflection point in the NIR at ${\rm log}\,\nu \sim
14.5$, which is thought to correspond to the sublimation temperature
of dust grains \citep{sanders89}, the IR hump at frequencies lower
than ${\rm log}\,\nu = 14.5$, which appears to be due to the
superposition of two blackbodies with different peak temperatures,
possibly related to emission from the hotter, inner edge and the
cooler, outer edge of the torus \citep{calderone12}, another gap in
the SED between the MIR and the radio band, and a radio upper limit
for this radio quiet source.

\begin{figure}
  \centering
  \includegraphics[width=0.48\textwidth]{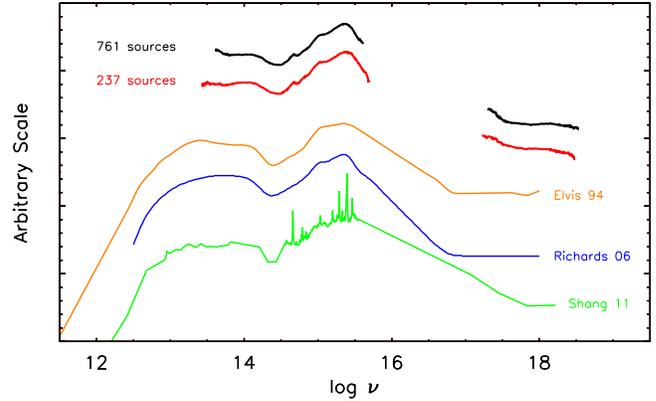}
  \caption{This figure compares the average SEDs produced in this work
    to average SEDs from the literature.  Plotted in black is the
    median SED produced from all 761 sources considered in this paper
    and plotted in red is the median SED of the 237 sources with well
    sampled SEDs which are make up the subsample of AGN considered in
    more detail in Section~\ref{section:variations}.  Plotted in
    orange is the mean SED of 29 RQQ from \citet{elvis94}, in blue is
    the mean SED of 259 SDSS quasars from \citet{richards06} and in
    green is the median SED of 27 RQQ from \citet{shang11}.  They are
    plotted on an arbitrary $y$ axis and have been artificially
    separated for clarity. }
  \label{fig:average_sed}
\end{figure}

We do not correct the photometry in any band for flux from broad
emission lines, but we note that such corrections would be small.  We
use the catalogue of quasar properties from \citet{shen11} to
determine typical equivalent widths of the main broad emission lines
in our sample and find the contribution to be $\leq 5$ per cent of the
flux in the SDSS $gri$ bands although this does increase to $\sim 10$
per cent for the broadest emission lines falling into the narrowest
SDSS $u$ band.  The average SED of our sample shown in
Fig.~\ref{fig:average_sed} does show a small feature at log
$\nu\sim14.7$ corresponding to the H$\alpha$ emission line which can
also be seen in some other SEDs throughout the paper (see in
particular Section~\ref{section:redshift}).  We do not see the effects
of any other broad emission lines in our SEDs.  Since we are unable to
correct for the H$\alpha$ emission line flux in higher redshift
sources, we do not attempt any corrections for line flux.  This small
feature in our SEDs does not significantly impact on our
interpretation of broadband changes in the SED shape.

Interpolation of the individual source SEDs is carried out using an
initial sampling of 20 points in a straight line between each pair of
consecutive luminosity estimates.  A second interpolation using 5000
$x$ points is then determined across the entire SED.  We carry out
this simple linear interpolation rather than attempting to fit a
complicated function (e.g. a quadratic) to avoid imposing a particular
shape on the SED, which may be unrepresentative.  The individual
errors on the luminosity estimates are not considered in the
interpolation, but are considerably smaller than the dispersion
observed in the average SED once created.  All luminosity estimates
flagged as upper-limits are visually inspected.  Any uninteresting
upper limits, in which the value quoted is higher than two real flux
measurements in bands either side, are excluded from the
interpolation.  29 interesting upper limits in which the value may be
tracing a real dip in the SED are included.

The bolometric luminosity of each source is determined by a simple
numerical integration according to Equation~\ref{eqn:lbol}, in which
the $L\sb{\nu}(i)$ values are taken from the 5000 interpolation
points and $\Delta\nu(i)$ is the width of each bin, unevenly
sized in linear space.  

\begin{equation}
  L\sb{\rm bol} = \int\limits\sb{0}\sp{+\infty} L\sb{\nu}\,d\nu \sim \sum\limits\sb{i}\sp{N} L\sb{\nu}(i)\,\Delta\nu(i)
  \label{eqn:lbol}
\end{equation}

In the interpolation we use a simple straight line (power law in
linear space) to interpolate across the EUV gap.  However, this may be
unrepresentative of the real SED shape in this frequency range leading
to an incorrect estimate of $L\sb{\rm bol}$.  In order to investigate
the size of this effect, we re-calculate the bolometric luminosity of
the sources using different interpolation shapes over this region.
The first, which gives an increased $L\sb{\rm bol}$ measurement
maintains the highest of the fluxes at the edge of the gap (UV or
X-ray) across the whole range, whilst the second maintains the lowest
of the fluxes across the gap and gives a lower $L\sb{\rm bol}$
measurement.  The average differences between these extreme estimates
and the estimate used in this analysis are $\lesssim 0.2$ dex for 76
per cent of the sources ($\lesssim 0.25$ dex for 88 per cent).

For 104 of the sources a radio detection is available and therefore
the SED interpolation, and hence $L\sb{\rm bol}$ determination,
includes the frequency range between \wise band 4 and radio.  For the
remaining 657 sources without a radio detection, this large frequency
range is not considered.  The difference between $L\sb{\rm bol}$
estimates, both including and excluding this range is $\lesssim 0.1$ dex
for 93 per cent of sources, making this effect much smaller than the
differences introduced by variations in the interpolation shape used
over the EUV gap.

\begin{figure}
  \centering
  \includegraphics[width=0.48\textwidth]{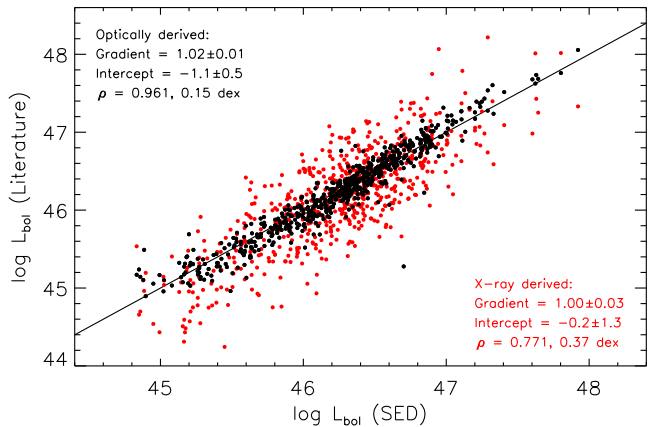}
  \caption{A comparison of the bolometric luminosity estimates
    obtained by integrating under our SEDs, with estimates from the
    literature.  Shown in black are optically derived estimates from
    \citet{shen08a} and shown in red are X-ray derived estimates using
    the original X-ray spectral fitting results from S11 and the
    bolometric correction factor from \citet{marconi04}.  Listed on
    the figure are the best-fitting linear trendline parameters, the
    Spearman rank correlation coefficient, $\rho$, and the level of
    scatter.  The solid black line is shown for guidance only and does
    not represent a fit to the data.}
  \label{fig:lbol_compare}
\end{figure}

Our $L\sb{\rm bol}$ estimates are in good agreement with literature
values.  Bolometric luminosities derived from optical monochromatic
luminosities and bolometric correction factors from \citet{richards06}
are given by \citet{shen08a}.  These values are plotted against our
estimates in Fig.~\ref{fig:lbol_compare} (black points) and show a
tight correlation and good agreement.  Estimates using the $2-10$ keV
luminosity from the original analysis in S11 and the luminosity
dependent bolometric correction of \citet{marconi04} are also shown in
Fig.~\ref{fig:lbol_compare} (red points).  These estimates show a
larger scatter, but are also in good agreement.  This similarity
between our \lbol estimates and those in the literature further
validates our SED interpolation technique which does not involve
complex model fitting.

We normalise each of the individual SEDs by the source's bolometric
luminosity.  We prefer this technique to the commonly used practice of
normalising the SED at a single frequency, which results in an
arbitrary `pivot point' and an artificial dispersion in the rest of
the SED.  We therefore plot $\textrm{log}(\nu L\sb{\nu}/L\sb{\rm
  bol})+C$ on the $y$ axis of figures, where the constant depends upon
the width of the bins used in the interpolation;
$\Delta\textrm{log}\nu=0.002$.  It has the value $C=-2.34$ implying
that each bin contributes $\sim 1/200\sp{\rm th}$
of $L\sb{\rm bol}$.  By normalising the individual AGN SEDs by their
bolometric luminosity, we expect the SEDs to have approximately the
same $y$-axis values.  This means that the figures are comparing the
\textit{relative} levels of emission at each frequency, rather than
absolute values and none of the dispersion is due to different
normalisations but is simply a result of different SED shapes.

Our sample includes luminous quasars with bolometric luminosities
greater than $7\times10\sp{44}\,\textrm{erg s}\sp{-1}$.  The possible
contamination from the host galaxies is therefore not expected to be
large.  As we have no direct observational data for the host galaxies,
nor any morphological information with which to correctly select an
appropriate host galaxy template, we can only estimate the galaxy
contribution.  We use equations (9)--(11) from \citet{elvis12} which
are adapted from \citet{marconi03} to include a redshift dependence,
to estimate the luminosity of the host galaxy in the NIR.  The
percentage of the total luminosity which can be attributed to the host
galaxy ranges from 0.1-1.3 per cent in the J and K bands to 0.1-1.5
per cent in the H band where the host contribution is known to be
largest.  However, this assumes that each AGN is accreting at the
Eddington luminosity, and whilst this is frequently used to determine
the minimum correction to be applied (e.g. \citealt{richards06}), here
we relax this assumption.  If each AGN is instead accreting at only 10
per cent of Eddington, we find that the host galaxy contributes up to
11 per cent in the H band.  This means that the host contibution can
move the SEDs we plot by at most 0.04 on the y axis, peaking in the
NIR H band, and decreasing rapidly at frequencies
$\textrm{log}\,\nu<14.0$ and $\textrm{log}\,\nu>14.6$.  This shift is
lower than the typical width of the SEDs we plot, however we exercise
caution when interpreting any differences in the SED shape at NIR
frequencies.

\begin{figure}
  \centering
  \includegraphics[width=0.48\textwidth]{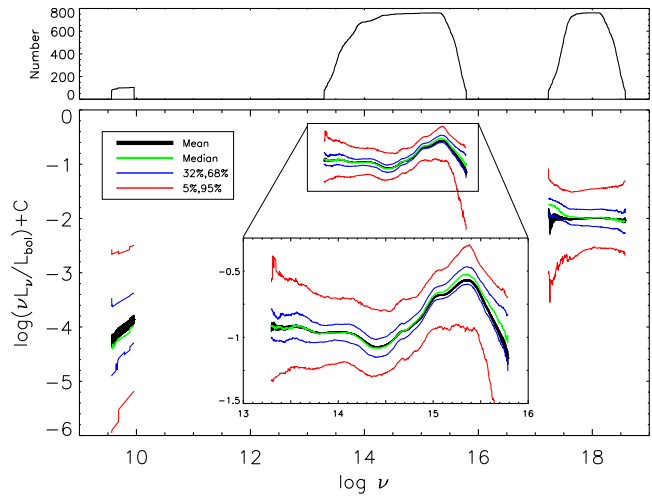}
  \caption{The average SED of the 761 type 1 AGN in the S11 sample.
    The median is shown by the green line, the $32\sp{\rm nd}$ and
    $68\sp{\rm th}$ percentiles are shown in blue, the $5\sp{\rm th}$
    and $95\sp{\rm th}$ percentiles are shown in red and the mean
    value is shown in black, where the thickness of the line
    corresponds to the error on the mean at each frequency point.  The
    large portions of the SED between radio and W4 and the FUV and
    very soft X-ray are omitted where they consist purely of
    interpolation points.  The ends of the SED are trimmed when the
    number of sources included falls to $<77$ (10 per cent of the
    initial sample size).  The number of sources included in the
    average SED determination at each frequency is shown in the top
    sub-panel.}
  \label{fig:mean_sed}
\end{figure}

In Fig.~\ref{fig:mean_sed} we plot the average SED of the 761 sources
in the sample.  On the $y$ axis we plot the median ${\rm log}(\nu
L\sb{\nu}/L\sb{\rm bol})+C$ value of the sources included in the
sample at that particular ${\rm log}\,\nu$ value (green line) along
with the $32\sp{\rm nd}$, $68\sp{\rm th}$, $5\sp{\rm th}$ and
$95\sp{\rm th}$ percentiles, which correspond to 1$\sigma$ and
2$\sigma$ error boundaries (blue and red lines).  Also shown in black
is the arithmetic mean of the SED values where the width of the line
corresponds to the standard error ($\alpha=\sigma/\sqrt{N}$) at that
frequency.  The mean gives a similar SED shape to the median in this
case, but using the median and percentiles is more effective at
rejecting outliers and gives a better representation of the dispersion
in the SEDs.

Between the radio and W4 band, and within the EUV gap, the SEDs
consist only of interpolation points and no real flux measurements
over a large frequency range.  Whilst these regions are considered in
the determination of $L\sb{\rm bol}$ to avoid an underestimate, we do
not plot them in the average SED.  The rest-frame frequency range
covered by each source's SED varies due to their redshift, so at
either end of the SED, and at the edges of the EUV gap, the number of
sources considered decreases from $N \rightarrow 0$ giving an
increased dispersion.  Since accurate median and percentile values
cannot be calculated from a low number of sources, we trim the
portions of the SED where the number of sources included at that
frequency falls below 10 per cent of the initial number included in
the subsample.  For subsamples with less than 50 sources, we trim the
SEDs when the number included falls below 5.

In order to give an accurate interpolation of the SED, it is important
to include as many real flux measurements as possible in order for the
SED to be well sampled across the full frequency range.  However, only
10 sources in the sample have a flux measurement (not including upper
limits) in each of the 19 wavebands.  This is mainly due to low
percentage coverages in the radio, FUV and NIR bands (see
Table~\ref{table:coverage}).  We therefore select a subsample of AGN
in which the majority of the SED is sampled with real flux
measurements.  We still require an FUV flux as this probes a
particularly important part of the SED and occurs at one end of the
high-energy SED gap.  However, we no longer require a radio, W3 or W4
flux.  The NIR region of the SED is important for mapping the
inflection point at ${\rm log}\,\nu \sim 14.5$, but this region has a
low coverage, only $\sim 60$ per cent.  We create an average SED of
196 sources which include each of the J, H and K bands (not
necessarily the Y band).  This is compared to a sample of 32 sources
which include at least one NIR measurement, (but not all 3 of J, H and
K), and another sample of 86 sources which do not include any.  The
SEDs which include at least some NIR data clearly show the inflection
point in the SED, but this feature is not apparent in the SED which
includes no NIR data.  This important feature appears to be missed if
no real flux measurements are included over this range.  However, we
note that the lack of a NIR inflection point may not simply be due to
data sampling alone and could be a real feature indicative of a
population of `hot-dust poor' type 1 AGN \citep{hao10}.

We create a final subsample including 237 AGN, 31 per cent of the
original sample.  They are required to include flux measurements in
each of the 3 X-ray bands, each of the 2 UV bands, all 5 optical
bands, at least one of the J, H or K bands, and the W1 and W2 MIR
bands.  Although some of the flux bands are not required, 83 per cent
of these sources do include all of JHK, 58 per cent include all of
YJHK, 96 per cent have a W3 flux, 71 per cent have a W4 flux and 16
per cent include a radio flux.  We also exclude sources whose
best-fitting X-ray spectral model in the original analysis of S11
required intrinsic absorption.  We identify 10 AGN in the sample with
possible dust reddening due to a relative colour $\Delta (g-i)>0.2$
after correction for Galactic reddening using an SMC extinction curve,
and a continuum shape of $\Delta (u-r) > \Delta (g-i) > \Delta (r-z)$
(See~\citealt{richards03}).  We do not exclude these AGN from our
final sample, but their presence does not impact our overall results.

The median SED of the final subsample is shown in
Fig.~\ref{fig:average_sed} plotted in red and is compared to the
median SED produced from all 761 sources in the original sample, shown
in black.  They are in good agreement suggesting that the results we
obtain from our good quality subsample are applicable to type 1 AGN in
general.  The reduced sample is slightly more biased towards lower $z$
and lower \lbol objects, but still covers a large range in these
parameters ($z: 0.11-3.29$, $\textrm{log}\,L\sb{\rm bol}:
44.85-48.10$).  In Fig.~\ref{fig:s11} we compare the \gmm and \lumx
distributions of the subsample of 237 AGN (red, dashed) to the full
sample of 761 AGN first reported in S11.  A KS test finds the \gmm
distributions not to be statistically different (significance
$=0.52$), but as noted above, the \lumx distributions are
significantly different (KS significance $=4.7\times 10\sp{-5}$).

Also shown in Fig.~\ref{fig:average_sed} are average SEDs from the
literature which have been reproduced here on an arbitrary scale, with
an artificial separation between them for clarity.  The mean SED of 29
X-ray bright, blue, RQQ from \citet{elvis94} is plotted in orange, the
mean SED of 259 mostly radio-quiet SDSS quasars from
\citet{richards06} is plotted in blue and the median SED of 27
optically bright RQQ at $z<0.5$ from \citet{shang11} is plotted in
green.  Our mean SED is in good agreement with these literature SEDs,
particularly in the optical/UV region where the `small blue bump' of
Balmer continuum and Fe\textsc{ii} line emission can be clearly seen.
It is a good match to the \citet{richards06} mean SED and the
underlying continuum of the \citet{shang11} SED.  Our SED shows
greater X-ray emission and lower MIR emission compared with the
\citet{richards06} SED.  This is related to the selection biases of
each sample as ours is biased against X-ray weak SDSS quasars and
theirs required a \textit{Spitzer} IRAC detection in all 4 MIR bands.

\section{SED variations}
\label{section:variations}
In this section we investigate variations in the average SED shape of
type 1 AGN using the 237 sources with SEDs which are well sampled over
a large frequency range.  We bin the sources according to their X-ray
spectral properties, redshift, their quasar sub-type, and physical
parameters.  The definitions of the further subsamples created for
this analysis are listed in Table~\ref{table:bins} along with the
number of sources included in each.  They are defined such that the
parameters are divided in a physically interesting way and where
possible the number of sources included in each bin is approximately
equal.  This ensures that any dispersion caused in the SED due to the
numbers of sources included is the same for each subsample.  The
average SEDs produced for these subsamples are shown in
Figs.~\ref{fig:x_var}, \ref{fig:obj_class} and \ref{fig:physical}.  In
each case the SEDs of the different subsamples are shown in different
colours.  The median value of ${\rm log}(\nu L\sb{\nu}/L\sb{\rm
  bol})+C$ is plotted, bounded by the $32\sp{\rm nd}$ and $68\sp{\rm
  th}$ percentiles, hence the width of each SED corresponds to a
$1\sigma$ error boundary.  The ends of the SEDs are trimmed when the
number of sources falls below 5, or 10 per cent of the initial number,
as described in Section~\ref{section:seds}.  The radio portion of the
SED is not shown in the majority of cases for clarity, but unless
stated otherwise it is consistent between the subsamples or there is
limited data available.  The main panel of each figure includes an
indicative bar showing the frequencies and regions of the SED which
correspond to each broad waveband range for a source at $z=0$.  In
Fig.~\ref{fig:physical} and the top panel of Fig.~\ref{fig:obj_class}
we also indicate the rest-frame frequency of the peak of the big blue
bump (BBB) with a $1\sigma$ error bar determined from the dispersion
of the individual BBB peaks of the included SEDs.  The peak frequency
we indicate is that of the peak in the median SED (which may not
directly correspond to the peaks as visible in the $68^{\rm th}$
percentile).  The top panel of each figure indicates how many
individual AGN have been used in the SED creation at that particular
frequency.

\begin{table}
 \centering
  \caption{The 237 sources which have a well sampled SED as described
    in Section~\ref{section:seds} are binned according to the
    different parameters listed below.  The definition of each of the
    sub-bins and the number of sources included in each are listed.
    Average SEDs binned according to these parameters are shown
    Figs.~\ref{fig:x_var}, \ref{fig:obj_class} and \ref{fig:physical}.
    \textit{Notes:} $^a$These sources are not included in the main
    sample considered in Section~\ref{section:variations}, but each
    has the same data requirements as the 237 sources which are
    included.}
  \label{table:bins}
  \begin{tabular}{lcc}
     \hline
     \multicolumn{1}{l}{Parameter} &
     \multicolumn{1}{c}{Definition} &
     \multicolumn{1}{c}{No. sources}\\
     \hline
     \gmm               & \:\:\:\:\:\:\:\:\:\:\:\:\gmm$<1.8$    &  78    \\[0.5ex]
                        & $1.8<$\:\gmm$<2.2$                    & 109    \\[0.5ex]
                        & $2.2<$\:\gmm\:\:\:\:\:\:\:\:\:\:\:    &  50    \\[1.5ex]

     Soft X-ray excess  & Power law                             & 195    \\[0.5ex]
                        & Power law + blackbody                 &  42    \\[1.5ex]

     X-ray absorption   & Power law                             & 237    \\[0.5ex]
                        & Absorbed power law                    &  12$^a$\\[1.5ex]

     Redshift           & $\:\:\:\:\:\:\:\:\:\:\:\:z<0.7$       &  97    \\[0.5ex]
                        & $0.7<z<1.2$                           &  67    \\[0.5ex]
                        & $1.2<z\:\:\:\:\:\:\:\:\:\:\:$         &  73    \\[1.5ex]

     Broad lines        & NLS1                                  &  11    \\[0.5ex]
                        & Non-NLS1                              &  94    \\[1.5ex]

     \rl                & RQQ                                   & 182    \\[0.5ex]
                        & RLQ                                   &  25    \\[1.5ex]

     \lumx       & \:\:\:\:\:\:\:\:\:\:\:\:\:\loglumx$<44.0$    &  69    \\[0.5ex]
                 & $44.0<$\:\loglumx$<44.5$                     &  99    \\[0.5ex]
                 & $44.5<$\:\loglumx\:\:\:\:\:\:\:\:\:\:\:      &  69    \\[1.5ex]

     \lbol       & \:\:\:\:\:\:\:\:\:\:\:\:\:\loglbol$<45.8$    &  80    \\[0.5ex]
                 & $45.8<$\:\loglbol$<46.3$                     &  77    \\[0.5ex]
                 & $46.3<$\:\loglbol\:\:\:\:\:\:\:\:\:\:\:      &  80    \\[1.5ex]

     \mbh        & \:\:\:\:\:\:\:\:\:\:\:\:\logmbh$<8.6$        &  58    \\[0.5ex]
                 & $8.6<$\:\logmbh$<9.0$                        &  67    \\[0.5ex]
                 & $9.0<$\:\logmbh\:\:\:\:\:\:\:\:\:\:\:        &  63    \\[1.5ex]

     \eddratio   & \:\:\:\:\:\:\:\:\:\:\:\:\:\:\eddratio$<-1.0$ &  64    \\[0.5ex]
                 & $-1.0<$\:\eddratio$<-0.7$                    &  71    \\[0.5ex]
                 & $-0.7<$\:\eddratio\:\:\:\:\:\:\:\:\:\:       &  53    \\[0.5ex]
     \hline
     \end{tabular}
\end{table}

\subsection{X-ray spectral parameters}
\label{section:x_var}
The X-ray spectral properties of the AGN in this sample were studied
in detail in S11.  A best-fitting X-ray power-law slope, $\Gamma$, was
determined for each and the presence of additional spectral components
such as a soft X-ray excess and intrinsic cold absorption were tested
for using an F-test at 99 per cent significance.  Since the X-ray
emission from AGN is thought to originate in the very central regions
close to the black hole, it provides a direct probe of the accretion
process.  Here, we investigate whether the properties of this emission
have an effect on the multiwavelength emission which is produced
further from the central source in the accretion disc and torus.

\begin{figure}
  \centering
      \includegraphics[width=0.49\textwidth]{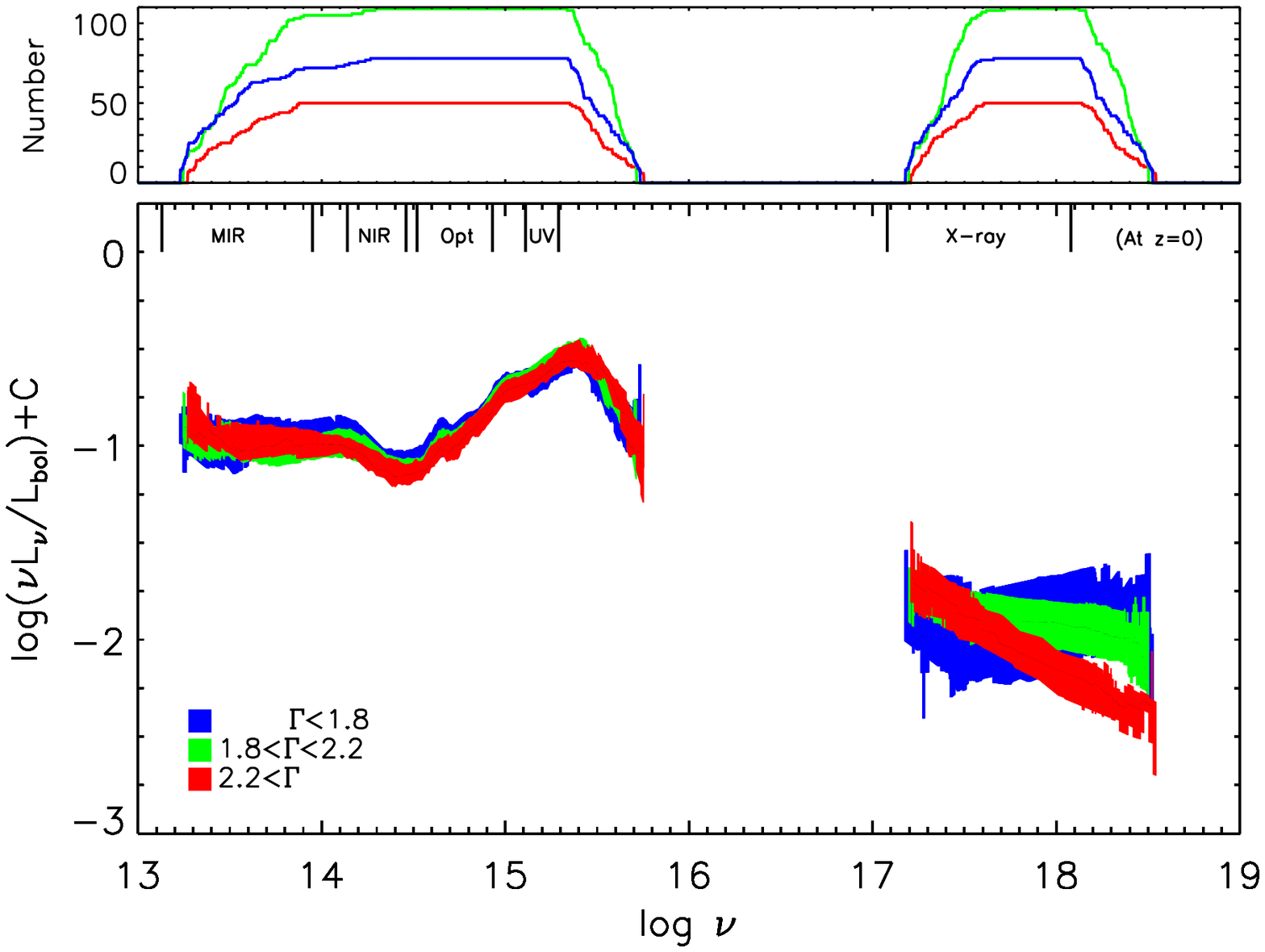}\\
      \vspace*{0.2cm}
      \includegraphics[width=0.49\textwidth]{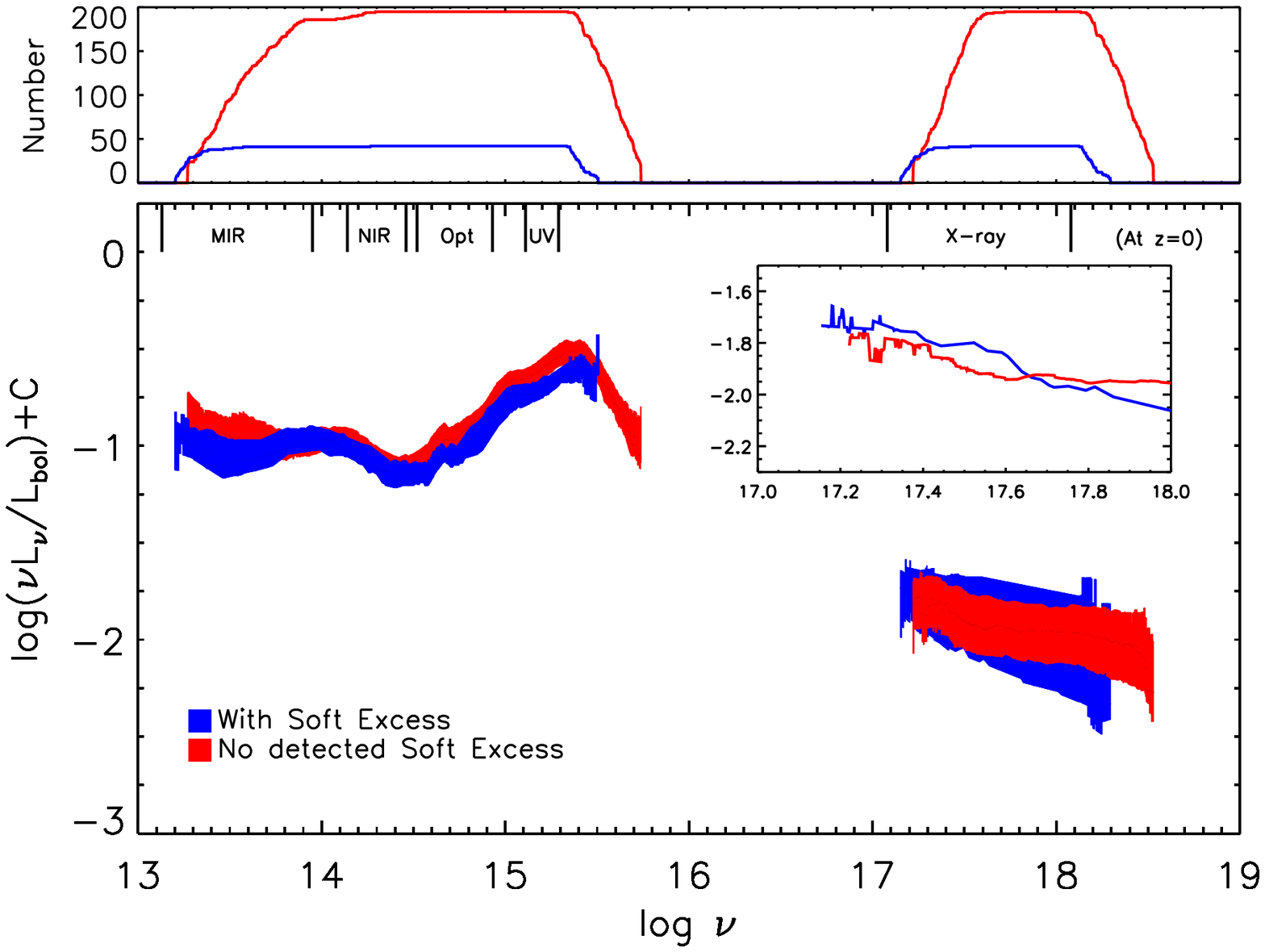}\\
      \vspace*{0.2cm}
      \includegraphics[width=0.49\textwidth]{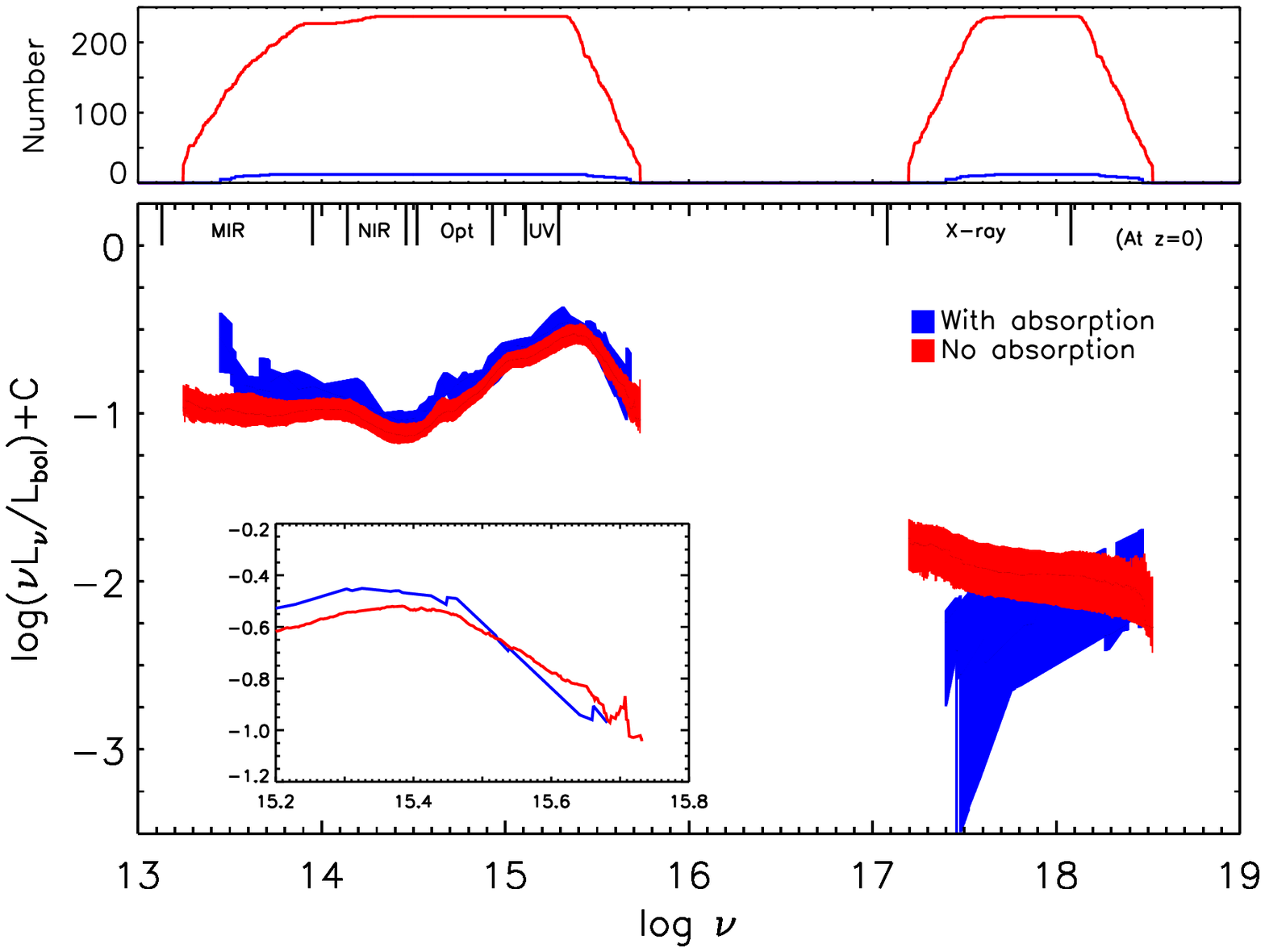}
  \caption{This figure shows average SEDs of sources binned according
    to their X-ray spectral properties.  The top plot shows variations
    with the X-ray power-law slope, $\Gamma$, the middle plot compares
    the SEDs of sources which did, or did not, significantly require a
    soft X-ray excess component in the original X-ray spectral
    modelling, and the bottom plot compares sources best-fit with a
    simple power-law model to those requiring significant X-ray
    absorption.  In this case the sources used are not included in
    the original sample of 237 sources, but do have SEDs which are
    well sampled with the same data requirements.  The inset panels
    plot the median values only.}
  \label{fig:x_var}
\end{figure}

Average SEDs of sources binned according to their best-fitting X-ray
power-law slope are shown in Fig.~\ref{fig:x_var} (top).  As expected
by definition, the SED shape in the X-ray band is different for each
of the subsamples, although the dispersion present in the flatter \gmm
SED is much larger than for the other subsamples.  This is due to the
presence of four sources with \gmm$<0.7$ giving a much larger range of
\gmm values in this subsample.  The SED shape at frequencies
$\textrm{log}\,\nu\sim 13-16$ is remarkably similar and the three SEDs
are almost entirely consistent within $1\sigma$ (there is some
deviation in the IR).  This suggests that the multiwavelength emission
of the sources has no dependence, or influence, on the shape of the
X-ray power-law emission.

Fig.~\ref{fig:x_var} (middle) compares the average SED of AGN in which
a soft X-ray excess component was significantly detected in their
X-ray spectra, to those which did not statistically require the
component in addition to a simple power law.  In general the two SEDs
have similar shapes suggesting the presence of a soft excess component
in the X-ray spectrum does not influence the source's emission at
other wavelengths.  This is expected, as the results of
\citet{scott12} showed that it is merely the X-ray spectral quality
which limits the detection of a soft excess and the feature is
ubiquitous in this sample of type 1 AGN.  The sources are hence from
the same overall population regardless of whether the soft excess is
significantly detected.  We therefore expect their SED shapes to be
the same and we continue to include these sources in our analysis.  We
might have expected to see a slight up-turn in the soft X-ray region
of the SED as a signature of the soft X-ray excess.  This feature is
indeed present in the median of the SED including significantly
detected soft excesses, but is masked by the reasonably large
dispersion on a relatively low number of sources in the subsample.  We
plot just the medians of the two SEDs in the inset panel.  This
up-turn is also visible in the original full sample of 761 sources in
which a soft excess was not necessarily significantly detected and can
be clearly seen in Fig.~\ref{fig:mean_sed}.  This further supports the
ubiquity of the soft excess.

The physical origin of the soft excess component is still debated in
the literature, but may be related to the intrinsic disc emission
(e.g. \citealt{done12}).  Unfortunately, in the average SED of sources
with a significantly detected soft excess, the BBB, which is tracing
the accretion disc, is poorly sampled.  This is because the soft
excess is easier to detect in lower $z$ sources as it lies in a
greater proportion of the \xmm EPIC instrument bandpass \citep{MOS,pn}
which introduces a strong redshift bias into the subsample.  However,
the peak of the BBB is sampled and is located at a lower relative flux
level ($\sim1\sigma$) and a higher rest-frame frequency for sources
with a significantly detected soft excess.

Fig.~\ref{fig:x_var} (bottom) compares the average SED of all 237 AGN
with well sampled SEDs to one made up of AGN with intrinsic absorption
present in their X-ray spectra.  Only 12 of such sources are used here
as we require the SEDs to include the same number of flux measurements
as the full sample.  The fluxes which make up the SEDs of the absorbed
sources are corrected for Galactic \nh (in the X-ray fluxes) and
Galactic \av (in the optical and UV fluxes), but are not corrected for
intrinsic absorption in any band.  The intrinsic X-ray absorption
column densities for these 12 sources range from $N\sb{\rm
  H}=(0.1-9.8)\times 10\sp{22}\,\textrm{cm}\sp{-2}$.

Despite the poorly defined shape of the average absorbed SED, which is
due to the low number of sources included, clear differences between
this and the unabsorbed SED are apparent.  As expected, there is a
lower level of emission at X-ray frequencies where the absorbing
material has the greatest effect with the SEDs being distinct at
$>1\sigma$ at frequencies $\textrm{log}\,\nu\,\le 17.5$.  However, the
optical/UV region of the SED does not display signs of similar
reddening.  This is partly due to the dispersion in the absorbed SED
masking the effect, which becomes visible in the inset panel where
only the median values of the two SEDs are plotted.  As this effect is
not large, it may suggest that the presence of gas absorption in the
X-ray regime does not necessarily imply that dust reddening in the
optical will also be present.  This topic is addressed further in
Scott et al. (in preparation) by studying the X-ray and optical
spectra of the sample in more detail.  There appears to be relatively
more emission at MIR frequencies, however we note that due to our
plotting method which normalises individual SEDs by their bolometric
luminosity, the significantly lower relative emission level at X-ray
frequencies will naturally result in an increased emission level in
the rest of the SED.  Even after taking this effect into consideration
the SEDs of absorbed and non-absorbed sources are distinct at
$>1\sigma$ at frequencies $\textrm{log}\,\nu\le 13.5$.  This may be
due to the absorbing material producing extra re-radiated emission.

Intrinsic X-ray absorption is expected to be present in up to 10 per
cent of optically classified type 1 AGN \citep{scott12} and we
consider such objects to belong to a different sub-population of
objects from their typical, unabsorbed counterparts.  This, and the
large difference between the absorbed and non-absorbed SEDs justifies
the exclusion of the absorbed sources from the subsample of sources
considered in our analysis.


\subsection{Redshift dependence}
\label{section:redshift}
Each of the sources in the sample has a spectroscopic redshift
measured by SDSS.  They cover the range $0.11 \le z \le 3.29$, with 95
per cent of sources at $z<2$, and both the mean and median redshift of
the sources is $z \sim 0.9$.  Here we investigate the effect of
redshift on the SED shape to determine whether an intrinsic,
non-evolving type 1 AGN SED exists.  There have been previous studies
investigating whether individual measures of the SED emission vary
with redshift.  In particular, there is thought to be no evolution of
\gmm with $z$
\citep{reeves00,piconcelli03,mateos05a,mateos05b,shemmer05,vignali05,tozzi06,just07,green09}
and no evolution of \aox with $z$
\citep{vignali03,steffen06,just07,green09,young09,stalin10}.  This
suggests that the emission at X-ray, UV and optical wavelengths is
non-evolving, and we should expect to find no shape evolution in the
SED at these frequencies.

\begin{figure}
  \centering
      \includegraphics[width=0.49\textwidth]{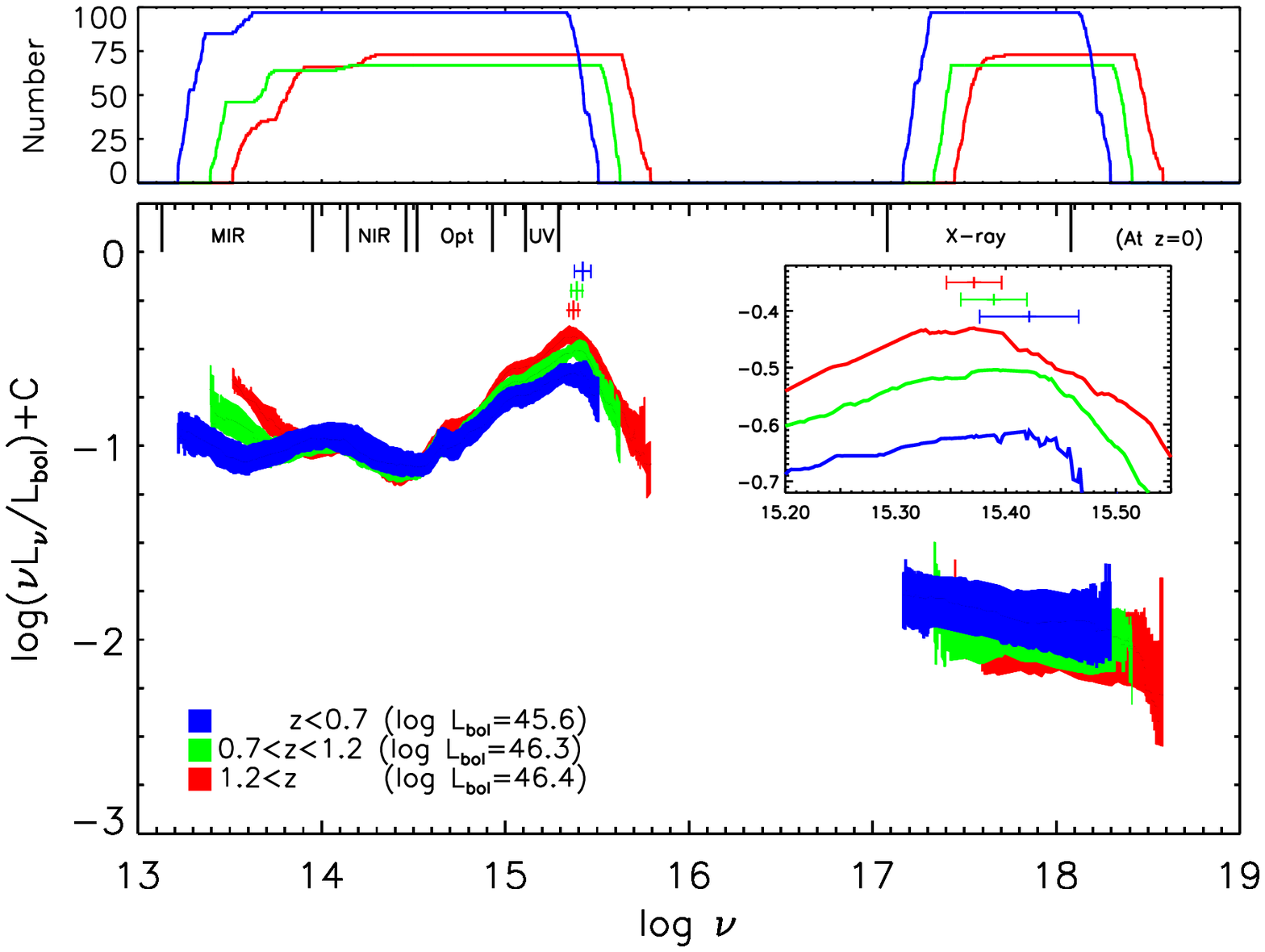}\\
      \vspace*{0.2cm}
      \includegraphics[width=0.49\textwidth]{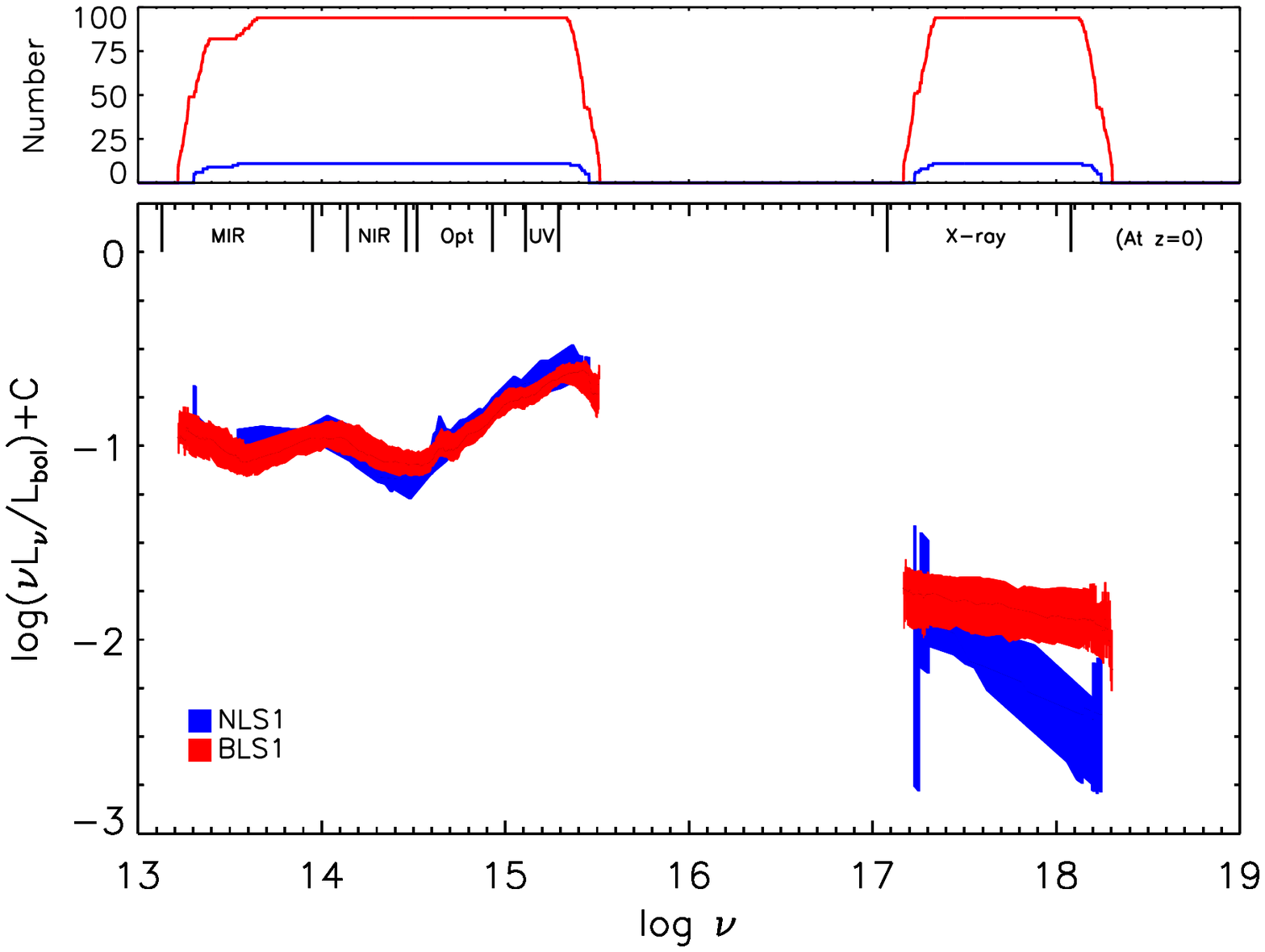}\\
      \vspace*{0.2cm}
      \includegraphics[width=0.48\textwidth]{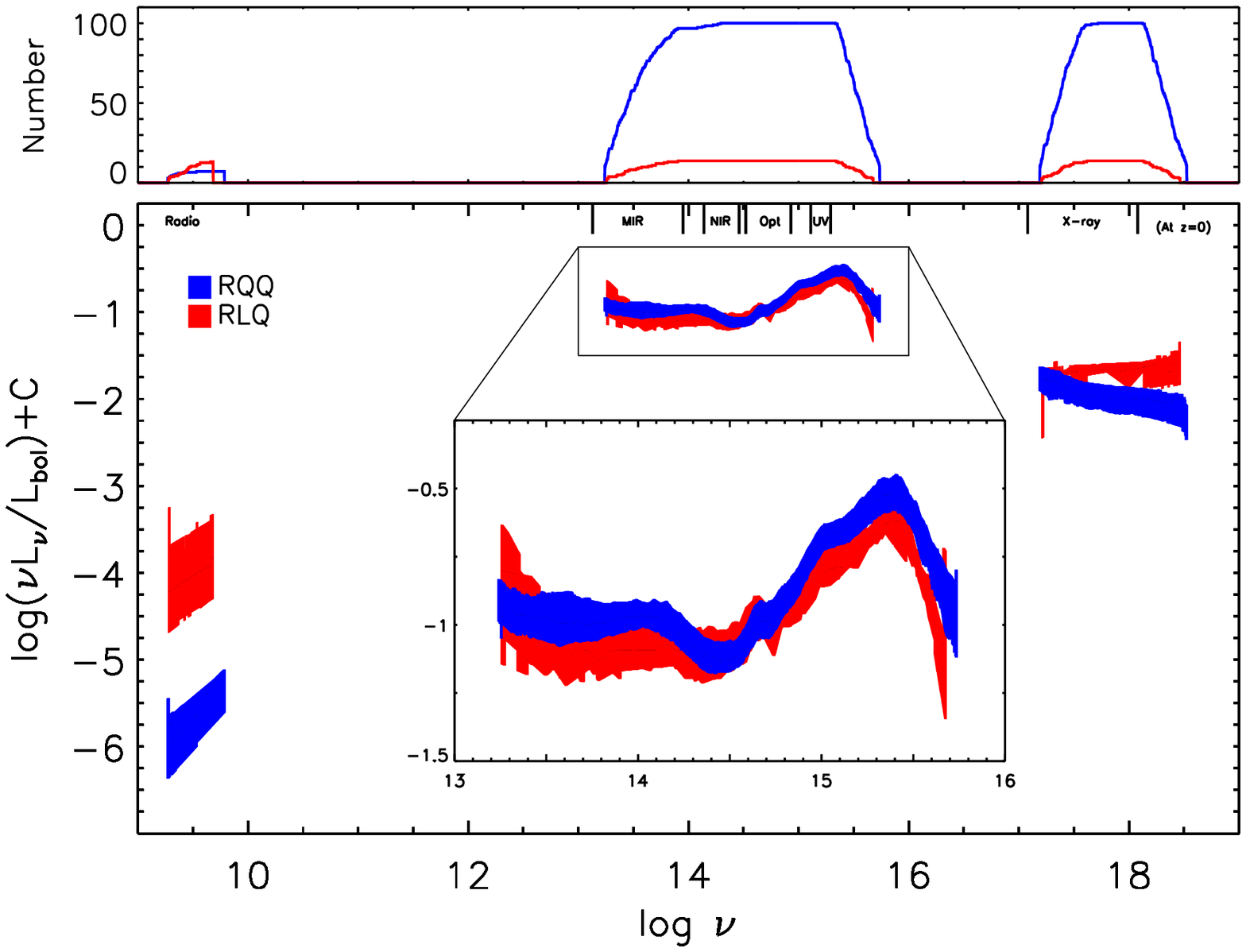}
  \caption{Top - average SEDs constructed from sources binned
    according to their redshift.  Listed in brackets is the median
    bolometric luminosity of the sources in each bin.  The coloured
    horizontal $1\sigma$ bars indicate the rest-frame peak frequency
    of the BBB in the median SED and the sub-panel plots the median
    SEDs only.  Middle - a comparison of the average SEDs of a
    subsample of NLS1 defined by the \citet{zhou06} catalogue and
    confirmed non-NLS1 sources.  Bottom - a comparison of the average
    SEDs of RQQ and RLQ.  Note the different axes scales on this
    figure.}
  \label{fig:obj_class}
\end{figure}

In higher redshift sources the observed fluxes correspond to higher
rest-frame frequencies and hence the average rest-frame SED lies at a
slightly different frequency range.  This effect is clearly visible in
Fig.~\ref{fig:obj_class} (top) and is particularly important in the
FUV region of the SED, which marks the beginning of the EUV gap.  The
Lyman continuum edge occurs at a rest-frame frequency of log 15.5
($=13.6\,{\rm eV}=912\,\textrm{\AA}$) and frequencies higher than this
are unmeasurable due to absorption by neutral hydrogen in our Galaxy.
However, for sources at $z\geq0.7$, the observed FUV flux has a
rest-frame frequency which probes into the EUV gap.  Therefore, in the
two average SEDs at higher redshift, some of this unpenetrable region
is being sampled.  These SEDs show a decrease in the UV, thought to be
probing the intrinsic disc emission.  In order to determine whether
our SEDs are affected by \lya absorption, we calculate the power-law
index, $\alpha$, where $F\sb{\nu}\propto\nu\sp{-\alpha}$ over the
range $1200\AA-2000\AA$ for each of the average SEDs.  The values are
consistent with the value given by \citet{shull12} for unabsorbed
sources, suggesting that no significant attenuation is present in our
SEDs.

The peak of the BBB occurs at a lower relative emission level in lower
redshift sources than for those at high $z$.  The relative emission
levels of the low $z$ and high $z$ SEDs are distinct at greater than
their $1\sigma$ dispersions from $\textrm{log}\,\nu\sim14.92$ until
the cut-off frequency at the edge of the EUV gap, however the SEDs are
consistent within their $2\sigma$ dispersions.  We quote the median
bolometric luminosity for each of the $z$ bins on
Fig.~\ref{fig:obj_class} (top) as the $L\sb{\rm bol}-z$ bias is
apparent.  It is therefore possible that the differences in the BBB
emission observed are a result of the changing luminosities of the
sources, rather than on $z$.  This is investigated in
Section~\ref{section:physical}.  The rest-frame frequency at which the
peak of the BBB occurs is also seen to shift to higher frequencies in
lower redshift sources as shown by the error bars plotted above the
SEDs.  However, we note that this shift is only $\sim 1\sigma$.

A striking difference between the SEDs is the higher relative emission
levels at MIR frequencies in the high $z$ sources.  The low $z$ and
high $z$ SEDs become distinct at $>1\sigma$ at a frequency of
$\textrm{log}\,\nu\sim13.75$, but are still consistent within
$2\sigma$ at the low frequency end of the SED.  This emission is
thought to be due to the re-radiation of higher frequency primary
emission from the central source, by dust further away.

The average SED made up of the lowest redshift sources shows a feature
at a rest-frame frequency of $\textrm{log}\,\nu \sim 14.7$.  This is
due to a number of sources at $z=0.4$ (the average for this subsample)
having an increased SDSS $z$ band flux.  This is likely due to the
broad emission line H$\alpha$ falling into this photometric band for
these sources.  This feature is not apparent in the mid-$z$ subsample
which has an average redshift of $z=1.0$ as H$\alpha$ falls into the
NIR J band and the sources considered are not required to include this
flux measurement.  H$\alpha$ falls in between NIR bands for sources in
the higher-$z$ subsample.

\subsection{Object class}
\label{section:obj_class}
A benefit of our large sample size is the opportunity to identify
reasonable numbers of AGN sub-types within the type 1 classification.
These include radio-loud sources which we identify using their radio
and optical fluxes, and narrow-line Seyfert 1s identified by the width
of their broad emission lines.

We identify 17 NLS1 through a cross-correlation of the S11 sample with
the \citet{zhou06} catalogue.  This includes 2011 NLS1 sources,
identified from SDSS DR3 within the range $z \leq 0.8$.  A well
sampled SED is available for 11 NLS1, the average of which is plotted
in Fig.~\ref{fig:obj_class} (middle) in blue.  This is compared to an
average SED of 94 sources which lie within the redshift range covered
by the catalogue and are hence confirmed as non-NLS1.

The average SED for the NLS1s shows a significantly steeper X-ray
slope than their broad-line counterparts, consistent with literature
results (e.g. \citealt{grupe10,jin12a}) and the relative X-ray flux
also appears to be lower in the average NLS1 SED.  However, the MIR-UV
SED shape does not appear to vary between the two types of source, in
that their $1\sigma$ widths are consistent throughout most of this
frequency range.  This suggests that the main physical difference
between them is in the X-ray emitting corona.  This may be a
difference in electron temperature or optical depth, since \gmm
depends on these parameters.  Results in the literature do report
changes in the SED at lower frequencies.  For example, \citet{jin12a}
find the peak of the BBB to be a more prominent feature in NLS1 with a
steeper $\Gamma$.  This is not apparent in this figure, possibly due
to the limited number of NLS1s.  However, in
Section~\ref{section:x_var}, we also found no change in the BBB with
variations in \gmm when we considered all type 1 AGN.

The radio loudness, $R\sb{\rm L}$, for each source was
determined using the radio flux from FIRST and the optical fluxes from
SDSS as described in more detail in S11.  A source was defined as
radio loud if \rl$>10$ \citep{kellermann89}.  Fig.~\ref{fig:obj_class}
(bottom) compares the average SED of RQQ to RLQ.  In order to display
the radio portions of the SEDs, both the $x$ and $y$ axes of this
figure have been extended.  In the case of the RQQ, the emission shown
at radio frequencies represents an upper limit because we only include
objects for which there was a significant radio detection by FIRST;
there are objects whose radio emission is too low to be significantly
detected and we do not include sources with radio upper limits in our
analysis.

As expected, the RLQ have greater amounts of radio emission than the
RQQ, with the SEDs being distinct at $>2\sigma$ over the entire radio
portion of the SED.  The RLQ also have more emission at X-ray
frequencies, the SEDs being distinct at $1\sigma$ over the frequency
range $\textrm{log}\,\nu=17.6-18.4$.  These observations are
consistent with the previous analysis in S11 and with observations in
the literature and suggest that the radio and X-ray emission is linked
(e.g. \citealt{zamorani81,worrall87,
  reeves00,brinkmann00,miller11,shang11}).  The emission at both
frequencies may be due to the presence of a jet, producing radio
photons through synchrotron emission, and X-ray photons from
Synchrotron Self Compton scattering.  The SED also shows a different
shape at X-ray frequencies.  This is again consistent with the results
of S11 in which \gmm was flatter in RLQ than RQQ.  This may be due to
contamination of the X-ray power-law emission by the jet.  The SED
shape in the MIR-UV region (shown enlarged) is very similar for both
the RQQ and RLQ, with the better defined RQQ SED lying almost entirely
within the $1\sigma$ error boundary of the RLQ SED (the BBB peak is
higher and there is more relative IR emission at $\textrm{log}\,\nu
\sim 13.7$).  The RLQ SED has a much larger dispersion due to the low
number of sources included, which also masks the inflection point in
the NIR.

These results imply that RQQ and RLQ are fundamentally the same
objects and hence produce the same broadband emission, but the
presence of a jet gives rise to the additional radio and X-ray
emission observed.  The similarity of the SEDs, with the exception of
the radio and X-ray bands has been observed previously
\citep{elvis94,shang11}.


\subsection{Physical parameters}
\label{section:physical}
In this section we investigate how the average SED shape changes with
variations in different physical parameters such as $L\sb{\rm X}$,
$L\sb{\rm bol}$, $M\sb{\rm BH}$ and $\lambda\sb{\rm Edd}$.  Finding
observational correlations between physical parameters can provide
constraints for theoretical models of the emission.


\begin{figure*}
  \centering
    \begin{tabular}{cc}
      \includegraphics[width=0.49\textwidth]{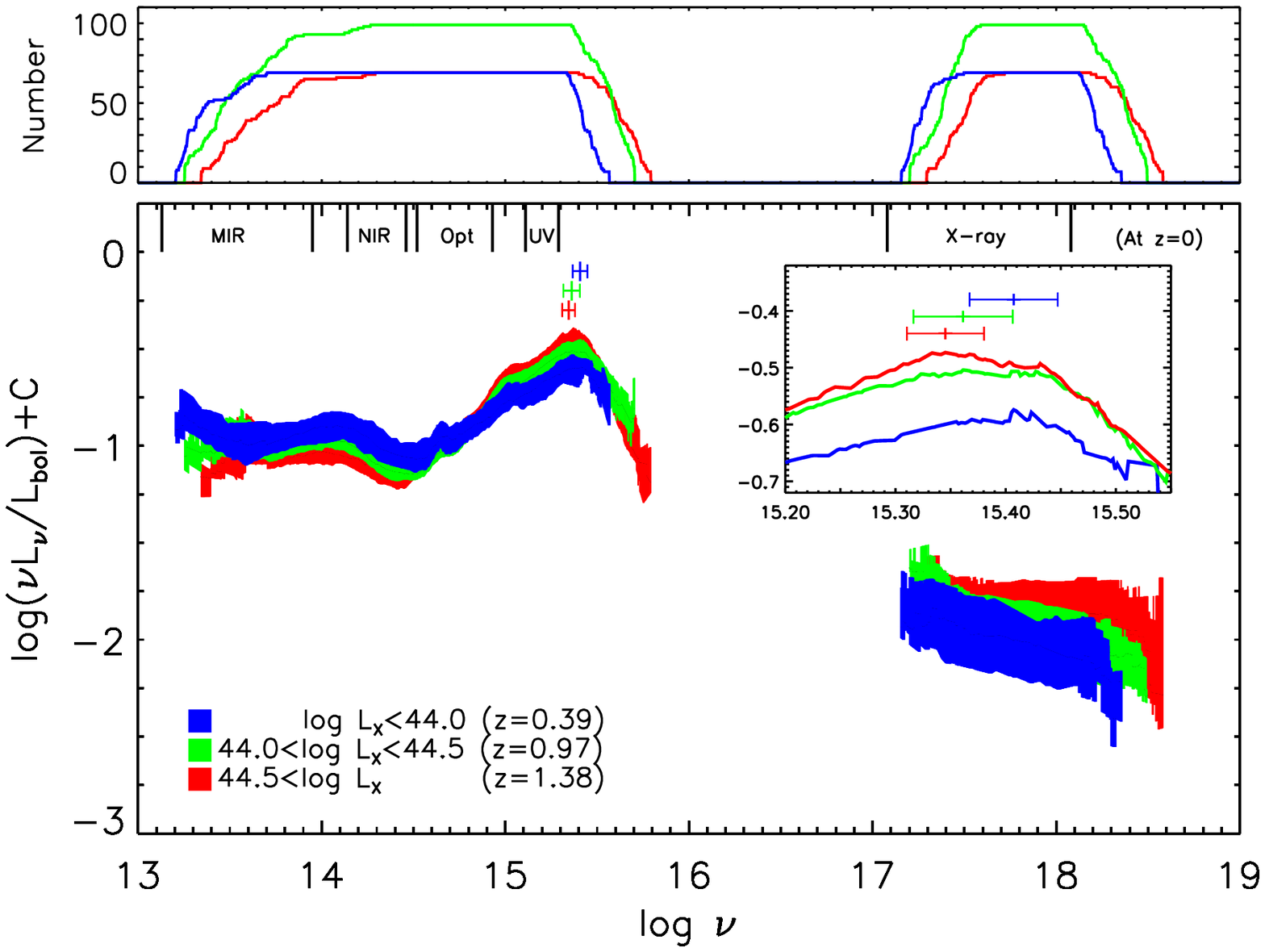} &
      \includegraphics[width=0.49\textwidth]{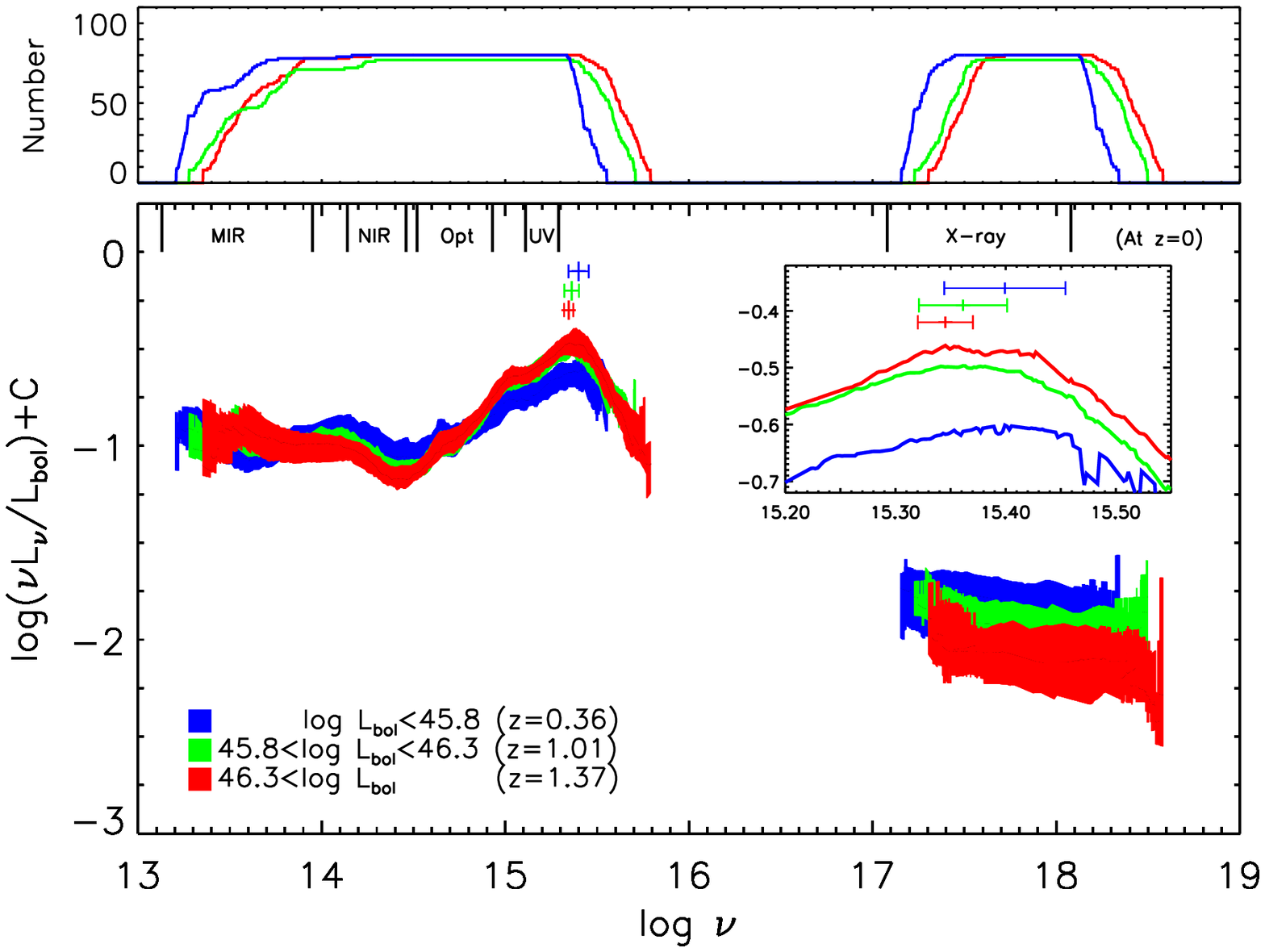} \\
    \end{tabular}
    \vspace*{0.2cm}
    \begin{tabular}{cc}
      \includegraphics[width=0.49\textwidth]{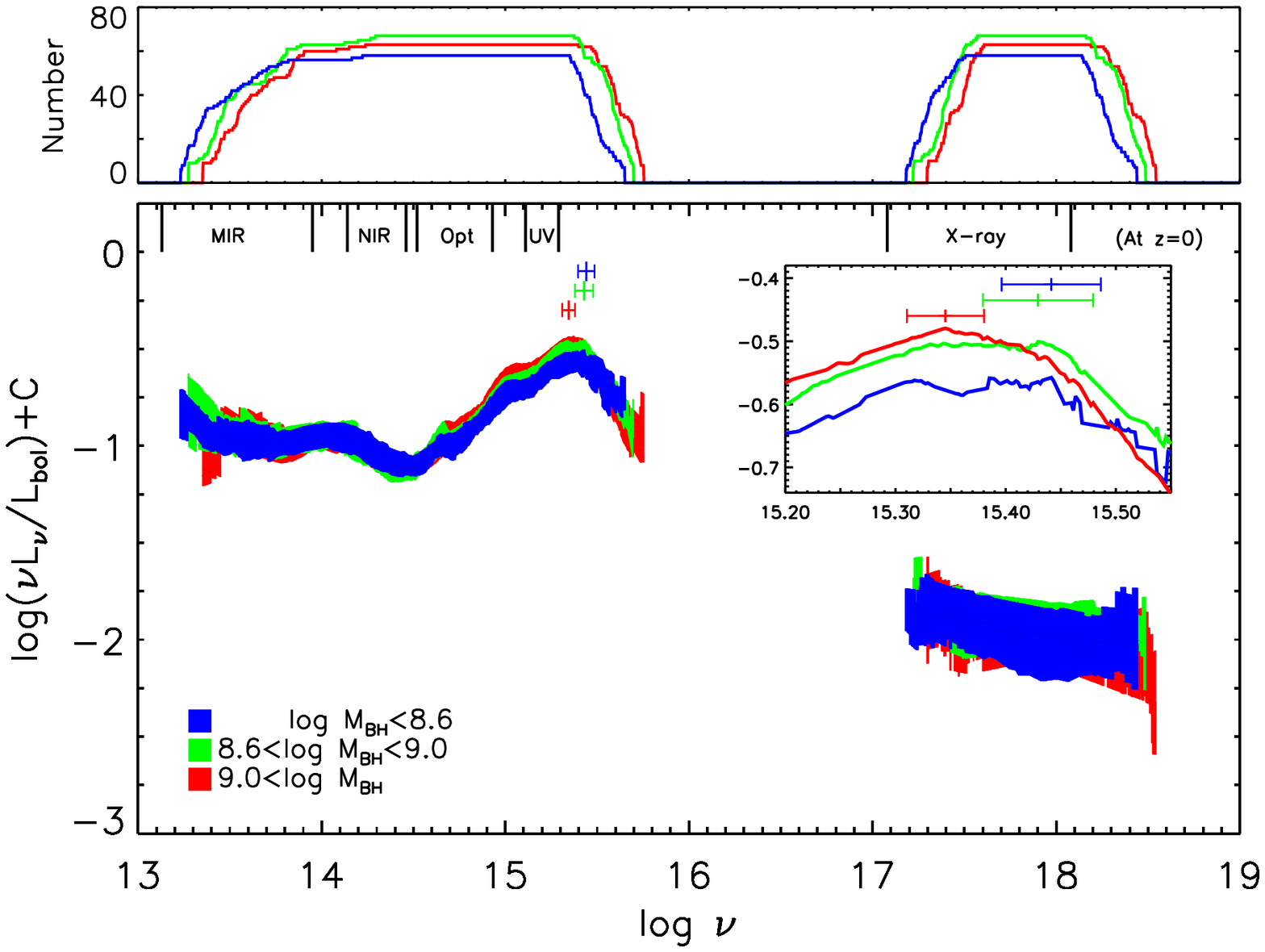} &
      \includegraphics[width=0.49\textwidth]{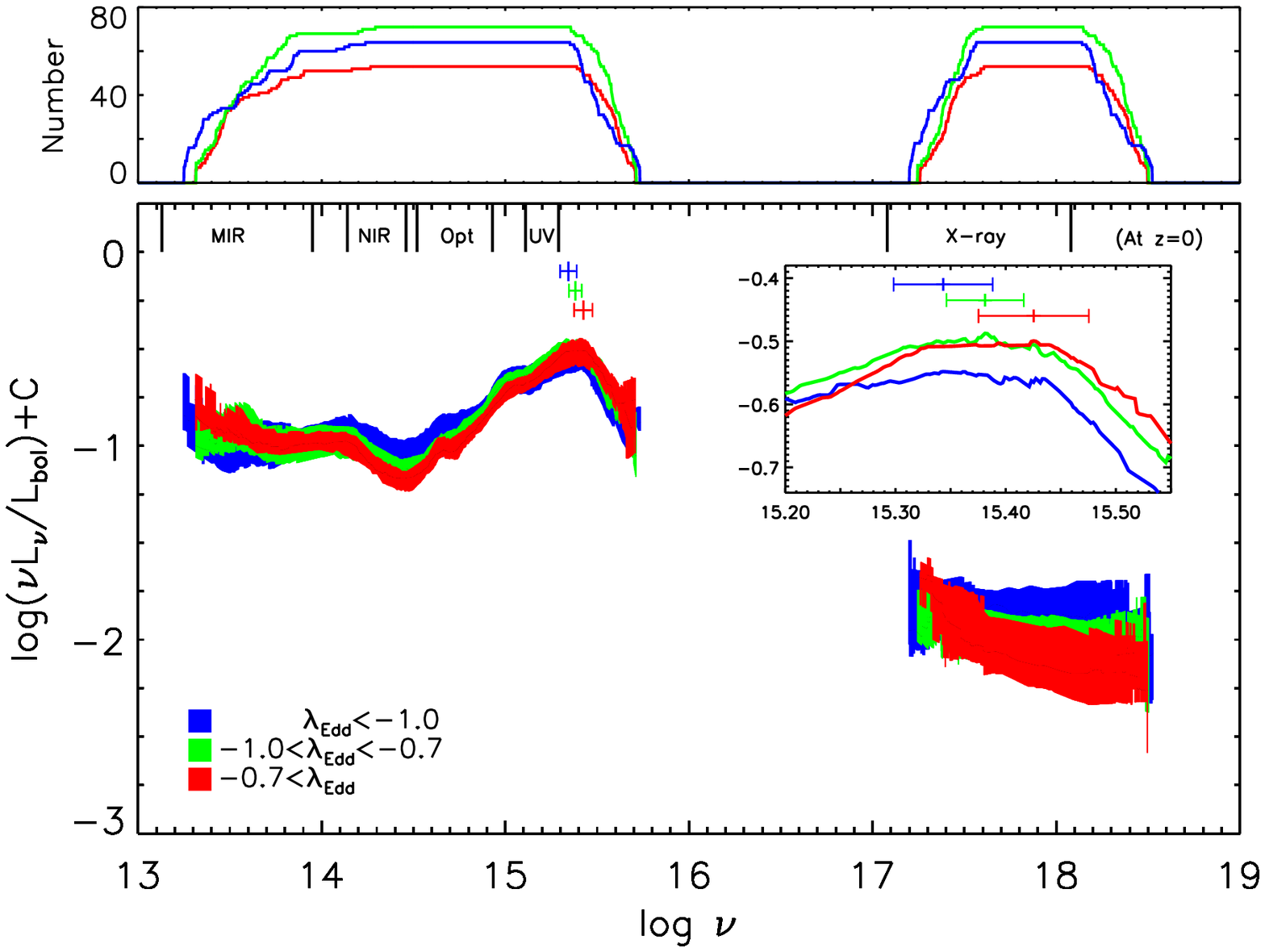}
    \end{tabular}
  \caption{This figure shows average SEDs of the 237 AGN which have a
    well sampled SED (see Section~\ref{section:seds} for details).
    The sources are further sub-divided based on the parameters
    $2-10\,{\rm keV}$ X-ray luminosity, \lumx (top left), bolometric
    luminosity, \lbol (top right), black hole mass, \mbh (bottom
    left), and Eddington ratio, \eddratio (bottom right).  The median
    redshift of the sources in each luminosity bin is shown in
    brackets on the top 2 figures.  In each figure the horizontal
    errors bars indicate the rest-frame frequency at which the peak of
    the BBB occurs in the median SED, with the $1\sigma$ subsample
    dispersion.  The inset panels show enlargements of the median SEDs
    around the BBB peak and these error bars.}
  \label{fig:physical}
\end{figure*}

Previous studies have investigated the luminosity dependence of
emission in individual bands (e.g. X-ray and optical:
\citealt{lusso10}; IR: \citealt{maiolino07}) and we therefore expect
to see some change in the SED shape with luminosity.
\citet{polletta07} presented average SEDs of type 1 AGN binned by
their $0.5-10$ keV luminosity.  No significant differences were seen,
but this is likely a result of the low numbers of sources included.
In Fig.~\ref{fig:physical} (top left) we show average SEDs produced
from 70--100 AGN in each 2--10\,keV X-ray luminosity bin.  It shows
that the X-ray contribution to the SED is approximately the same in
each subsample, in that they are consistent with each other within
their $1\sigma$ errors shown by the width of the plotted SED.
However, the median is higher in the highest \lumx subsample
suggesting a slightly larger X-ray contribution, although this is not
statistically significant.  This would agree with the requirement of a
luminosity-dependent X-ray bolometric correction
(e.g. \citealt{marconi04}).  The amount of emission contributed from
each band within the MIR-UV frequency range does vary between
subsamples, although the combined contribution must be approximately
the same.  The average SED made from low \lumx sources has a rather
flat shape, whereas the high \lumx sources show a higher BBB peak and
lower IR emission (the SEDs are distinct at $>1\sigma$ between
$\textrm{log}\,\nu \sim 14.1-14.4$ and $14.9-15.4$).  This suggests
that in sources with higher X-ray luminosities, a greater amount of
the AGN's emission comes out in the optical/UV.  This is in agreement
with the anti-correlation found between \aox and the optical
luminosity, $\textrm{log}\,l\sb{2500\angstrom}$ implying a non-linear
relationship between the optical and X-ray emission
\citep{avni86,vignali03,strateva05,steffen06,just07,vasudevan07,shemmer08,young09,stalin10,lusso10}.
These SEDs show a significant difference at radio frequencies with
relatively more radio emission in sources with higher X-ray
luminosities, consistent with the results of the RLQ in
Section~\ref{section:obj_class}.

Fig.~\ref{fig:physical} (top right) shows how the SED shape varies
with changes in the bolometric luminosity of the source.  The
$L\sb{\rm bol}-z$ bias is apparent in this figure with the highest
\lbol SED covering higher frequencies due to the higher-$z$ sources it
includes; the median $z$ of the low \lbol SED is $0.36$,
$\bar{z}=1.01$ for the mid \lbol SED and $\bar{z}=1.37$ for the high
\lbol SED.  The main difference between the SEDs is the steeper
gradient in the higher luminosity SED between the NIR inflection point
and the peak of the BBB.  This is similar to the behaviour of the SEDs
when binned according to $L\sb{\rm X}$.  As discussed in
Section~\ref{section:seds}, the difference in the NIR may be an
indication of some contamination by the host galaxy which is expected
to have the largest effect in the H band for sources at $z=0.5$
\citep{mcleod95}.  In the lowest luminosity sources the inflection
point appears less pronounced, possibly due to the addition of
emission from the host galaxy, whose SED has the opposite shape.

As the SEDs are normalised by their bolometric luminosity, the $y$
axis indicates the relative amount of emission at each frequency.
This is similar to the way in which bolometric correction factors
would quantify this.  In the X-ray band, the amount of emission
appears to vary with the bolometric luminosity of the source with
X-rays contributing more in the low luminosity SED and less in the
high luminosity SED.  This implies that a luminosity dependent X-ray
correction is required, supporting the results of \citet{marconi04}.
However the low and high \lbol SEDs are only distinct in the X-ray
band at $\le1\sigma$.  The high luminosity sources also show
relatively more emission in the BBB ($>1\sigma$) implying that the
bolometric correction factor for these frequencies must also have a
luminosity dependence.  However, at IR frequencies
$\textrm{log}\,\nu\lesssim 14.4$, the SEDs are almost consistent
within their $1\sigma$ dispersions suggesting that within the
luminosity range of sources included by our sample, the bolometric
correction factor in these bands is constant and is not dependent on
the luminosity of the source.  This is also shown in
\citet{hopkins07}.

Virially determined black hole mass estimates taken from
\citet{shen08a} are available for 188 (79 per cent) of the sources.
The average SEDs binned according to this parameter are shown in
Fig~\ref{fig:physical} (bottom left).  The three SEDs show a
remarkably similar shape at all frequencies (all 3 are consistent
within their $1\sigma$ dispersion), suggesting that the
multiwavelength emission from type 1 AGN is largely insensitive to the
mass of the accreting black hole.  However, the peak of the BBB occurs
at a higher relative flux level and lower rest-frame frequency for the
SED containing AGN with higher mass black holes.  A correlation
between \gmm and \mbh has been found in the literature
\citep{porquet04,piconcelli05,kelly08,risaliti09}, but is not apparent
in this figure.  However, it is generally thought that the underlying
driver of this correlation is one with accretion rate
\citep{wang04,porquet04,piconcelli05,bian05,shemmer06,shemmer08,risaliti09,grupe10}.

We therefore consider whether the SED shape varies with accretion rate
for which we use the Eddington ratio, defined here as $\lambda\sb{\rm
  Edd}\,=\textrm{log}\,(L\sb{\rm bol}/L\sb{\rm Edd})$, as a proxy.  We
use the bolometric luminosity determined from the SED fitting
described in Section~\ref{section:seds} and the black hole mass
estimates from \citet{shen08a} to determine the Eddington luminosity.
Fig.~\ref{fig:physical} (bottom right) shows that the X-ray emission
is flatter in the SED made up of low Eddington ratio sources, and
steeper in the higher Eddington ratio SED.  The MIR-UV region is
broadly similar for each of the subsamples, however, the gradient
between the inflection point and the peak of the BBB is steeper in the
higher Eddington ratio SED.  This is mostly due to the relatively
lower emission in the NIR ($<1\sigma$), but is also partly due to a
higher optical/UV peak.  An increased disc emission in higher
Eddington ratio sources was also observed by \cite{jin12c}.  This
figure also suggests that the peak of the BBB may be shifted to higher
frequencies with increasing Eddington ratio, albeit not with high
significance ($\sim 2\sigma$).


\section{Discussion}
\label{section:discussion}
In Section~\ref{section:variations} we presented average SEDs of
different subsamples of type 1 AGN and found their shapes to vary.
Such variation suggests that the physical structure or emission
mechanisms within the individual AGN may vary, and it is important to
understand the physical drivers behind these changes.  Some of the
more dramatic differences we see are due to the sources belonging to
different sub-populations.  X-ray absorbed type 1 AGN show a large
deficit of emission at optical to X-ray frequencies and an excess of
emission at MIR frequencies.  As their SEDs are so different to the
unabsorbed sources, we regarded them as a separate population of
objects and excluded them from the rest of the study.  Radio-loud
quasars show a very similar SED to radio-quiet sources over the IR-UV
frequency range, but show considerable differences at radio and X-ray
frequencies, likely related to an additional jet component.  NLS1 also
have a very similar SED shape to their broad-line counterparts over
the majority of frequencies but show large differences in the X-ray
regime suggesting different conditions in the X-ray emitting corona.
These results suggest that the sub-populations are all fundamentally
the same type of object, but have some physical differences which can
cause large variations in their SEDs.  When we consider changes with
underlying physical parameters, we see more subtle changes in the SED
shape.

The X-ray emission properties of AGN are critical for the
interpretation of the accretion process and it is important to
determine how this emission affects the rest of the source.  We found
that the X-ray spectral parameters appear to have little influence, or
dependence, on the emission in the rest of the SED.  Therefore the
X-ray emission is a poor indicator of the emission properties at other
frequencies.  We see a wide range in the allowed spectral properties
at X-ray wavelengths, possibly due to their greater sensitivity to
physical changes in the very central regions of the AGN close to the
black hole.  In contrast, the emission from the accretion disc sitting
at larger radii shows very similar spectral properties for all AGN.
This suggests that the X-rays, and changes in this emission, do not
have any effect on the accretion disc properties.

The BBB, which traces the accretion disc, is one of the major features
in the SEDs.  In our analysis we have seen subtle changes in both the
relative flux level of the peak of this component, and the rest-frame
frequency at which this peak occurs.  If the accretion disc is assumed
to radiate locally as a blackbody, this peak frequency can be used to
determine the inner disc temperature.  This depends on both the mass
of the black hole and the accretion rate (see Equation~\ref{eqn:tin})
and for a typical AGN with a $10\sp{8}\msol$ black hole and
$\dot{m}=0.1$, is $T\sb{\rm in}\sim 200\,000$ K ($\sim 0.02$ keV).

\begin{equation}
  T\sb{\rm in} \propto \left(\frac{\mdot}{\mdot\sb{\rm Edd}}\right)\sp{1/4} \left(\frac{M}{\msol}\right)\sp{-1/4}
  \label{eqn:tin}
\end{equation}

For lower redshift AGN the peak of the BBB lies at a higher rest-frame
frequency and a lower relative flux level than for those at higher
$z$.  This is the same behaviour displayed by AGN which include a
significantly detected soft X-ray excess component, as they are
generally low $z$ sources due to the bias involved in the detection of
the component.  The shift in BBB peak frequency seen between low and
high redshift SEDs corresponds to a $\sim 10$ per cent decrease in the
inner disc temperature ($\Delta T \sim 22000\,\textrm{K}$).  This
could be due to either higher black hole masses or lower accretion
rates in the high $z$ AGN.  The peak frequency of the BBB does not
change with luminosity (\lumx or $L\sb{\rm bol}$), but the relative
flux level does change, indicating that the amount the BBB emission
contributes to the total luminosity of the source increases with
luminosity.  The relative flux level of the peak is also higher for
AGN with higher $M\sb{\rm BH}$ and occurs at a lower rest-frame
frequency.  We also find that the BBB peak frequency shifts with
Eddington ratio, giving lower temperatures in lower accretion rate
sources.  The peak frequency, expressed here in energy, shifts from
$\sim 0.009$ keV in low accretion rate AGN to $\sim 0.01$ keV in the
mid and high Eddington ratio bins, which show approximately the same
peak frequency and flux.  We expect the inner disc temperature to
depend on both the black hole mass and the accretion rate as indicated
by Equation~\ref{eqn:tin}, but the accretion rate itself also depends
on the mass of the black hole.  It is therefore unclear whether the
main driver of the peak frequency shift is that of the black hole mass
or the accretion rate.  However, we also observe some slight changes
in the X-ray emission in the SEDs binned by Eddington ratio,
specifically flatter power-law slopes in lower accretion rate sources.
When coupled with the observed shift in the peak frequency of the BBB
we can draw analogies between the behaviour observed here in AGN and
the different spectral states observed in Galactic black hole binaries
(e.g. \citealt{belloni10}).  In the high/soft state in which the
accretion rate is high ($-1<$\,\eddratio$<-0.3$) and the X-ray
power-law slope is steep, the disc is thought to extend close in to
the black hole resulting in a hotter inner disc temperature ($T\propto
r\sp{-3/4}$).  Conversely in the low/hard state, where the accretion
rate is low (\eddratio$<-1$) and the X-ray power law is flat, the
inner accretion disc is thought to truncate further out from the black
hole, leaving a geometrically thick, optically thin
advection-dominated accretion flow, and a lower inner disc temperature
\citep{esin97}.  This scenario is consistent with the results shown by
the different average SEDs presented here, in which the lowest
Eddington ratio subsample includes objects in the low/hard state,
whilst the top two bins fall into the high/soft state.

We include \wise data in our SEDs which samples the MIR region probing
dust, allowing us to investigate the physical properties of this
material.  We see clear differences in the MIR emission of absorbed
sources and the SEDs produced from sources binned by their redshift.
There is relatively more MIR emission in the higher $z$ sources than
the low $z$ sources, which due to the flux limited sample may be a
luminosity effect instead.  However, when we bin the sources according
to their bolometric luminosities, no difference is seen.
\citet{calderone12} fit the MIR emission of a similar sample of AGN
with a superposition of two blackbodies with temperatures of 1500 K
and 300 K, possibly representing the inner (hotter) and outer (cooler)
edge of the dusty torus.  In Fig.~\ref{fig:obj_class} (top; $z$
dependence), we see a possible blackbody peak at a (log) frequency of
14.1, similar to \citet{calderone12}.  This appears to be the same for
each of the three redshift bins.  We also see evidence for another
blackbody component at lower frequencies, although we are only
sampling the tail of the distribution rather than the peak.  However,
we do see that the peak temperature would be higher in the highest $z$
SED, implying a higher temperature of dust for these sources.  If
these two temperatures were to correspond to the inner and outer edges
of the dusty torus, it is unclear why only one of these temperatures
should show a $z$ dependence.  We suggest that the two components may
instead be sampling different regions of dust.  In the case of the
absorbed sources, we estimate that $\sim 10$ per cent of \lbol is
absorbed in the X-ray and optical region, with $\sim 7$ per cent of
this being re-emitted in the MIR.  This additional MIR emission
corresponds to dust with a temperature of $\sim 1000$ K.

This work has the advantage of studying a large sample of SEDs which
include AGN with a wide range of physical and spectral properties.
However, the sample was created from the optical SDSS quasar catalogue
and the X-ray source catalogue 2XMMi, along with additional archival
multiwavelength catalogues which imposes some biases and caveats into
the resulting sample.  It is biased against the weakest X-ray sources
as only AGN with $>75$ X-ray spectral counts are included, and the
sensitivity limit varies across each of the frequencies we sample.
However, in this work it is more important to accurately sample the
shape of the SED by including as many flux measurements as possible
than to have a flux complete sample of objects.  We therefore selected
a subsample of AGN with detections in as many flux bands as possible.
By requiring the inclusion of NIR and UV data in our SEDs we do create
a slight bias towards lower redshift and lower luminosity sources, but
still cover a wide range in these parameters.  We do not include any
data at FIR frequencies leading to a large interpolated region between
the W4 and radio bands.  However, the majority of the emission at
these frequencies would be due to starburst activity rather than the
AGN \citep{netzer07,mullaney12}.

The multiwavelength data we use are from non-contemporaneous
observations where in some cases the data was taken many years apart.
As AGN are known to be variable, this could mean that some of our SEDs
are constructed from data points taken when the intrinsic SED shape of
the source was different.  Spectral changes seen at optical and X-ray
frequencies are more rapid than the variability timescales in the IR
band.  However, in this work we focus on investigating variations in
average SEDs created from a large number of individual objects which
makes any such temporal variations less important.


\section{Conclusions}
\label{section:conclusions}
In this paper we have created broadband SEDs of a large sample of type
1 AGN using archival data from \xmm (X-ray), \galex (UV), SDSS
(optical), UKIDSS (NIR), 2MASS (NIR), \wise (MIR) and FIRST
(radio). 237 AGN (31 per cent) of the initial sample have well sampled
SEDs over a large frequency range from which we created average SEDs,
binned according to different properties and physical parameters.  We
investigated variations in the shape of these SEDs in order to
determine how changes in these parameters affect the AGN emission at
different frequencies.

We found that the SEDs of type 1 AGN sub-populations do show
considerable differences.  X-ray absorbed sources show a large deficit
in the optical/X-ray regime and an excess at MIR frequencies,
consistent with re-emission from the same absorbing material.  RLQ are
consistent with being the same type of physical source as RQQ, but
with an additional jet component which contributes extra X-ray and
radio emission from non-thermal processes.  NLS1 show significant
differences in the spectral shape of their X-ray emission, perhaps
related to physical variations in the corona, but the rest of the SED
is similar to their broad-line counterparts.
 
The multiwavelength emission does not appear to have any dependence on
the shape of the X-ray spectral emission. We see a wide variety of
X-ray spectral slopes, but only a narrow range of IR-UV variations.
This indicates that although the accretion processes occurring in the
central regions close to the black hole can vary substantially, the
emission from the accretion disc at large scales and the torus are
generally similar for all AGN.  Similarly the presence of a soft
excess does not alter the rest of the SED emission.  We also found
that the X-ray emission contributes a lower amount to the overall
emission in AGN with high bolometric luminosities.  This explains, and
illustrates, the need for a luminosity-dependent X-ray bolometric
correction in which the X-ray contribution decreases with increasing
$L\sb{\rm bol}$.  AGN with higher X-ray luminosities also emit a
greater amount of their emission at optical/UV frequencies which
agrees with the previously known anti-correlation between \aox and
$l\sb{\rm opt}$.

We see some shifts in the peak frequency of the BBB which suggest a
change in the inner disc temperature.  The shift seen between low and
high $z$ sources is likely to be a consequence of an underlying change
in accretion rate or black hole mass.  Increased inner disc
temperatures in higher accretion rate sources may be suggestive of a
change in the accretion disc structure in analogy with spectral states
observed in Galactic black hole binaries.

Our SEDs come from type 1 AGN spanning a wide range of parameters.
Given this, it is perhaps remarkable that the diversity in the SEDs is
relatively small, particularly at frequencies lower than X-rays and
higher than radio.  Subtle dependencies on a number of physical
properties are evident, but there is no apparent single driver for
these variations.


\section*{Acknowledgments}
We thank the referee for detailed comments on the manuscript.  We also
acknowledge financial support from the UK Science \& Technology
Facilities Council.  This work is based on observations obtained with
\textit{XMM-Newton}, an ESA science mission with instruments and
contributions directly funded by ESA Member States and NASA.  It also
uses data products from: the Wide-field Infrared Survey Explorer,
which is a joint project of the University of California, Los Angeles,
and the Jet Propulsion Laboratory/California Institute of Technology,
funded by the National Aeronautics and Space Administration; the Two
Micron All Sky Survey, which is a joint project of the University of
Massachusetts and the Infrared Processing and Analysis
Center/California Institute of Technology, funded by the National
Aeronautics and Space Administration and the National Science
Foundation; the NASA Galaxy Evolution Explorer which is operated for
NASA by the California Institute of Technology under NASA contract
NAS5-98034; the UKIRT Infrared Deep Sky Survey (UKIDSS); the VLA FIRST
Survey; and SDSS, funding for which has been provided by the Alfred
P. Sloan Foundation, the Participating Institutions, the National
Science Foundation, the U.S. Department of Energy, the National
Aeronautics and Space Administration, the Japanese Monbukagakusho, the
Max Planck Society, and the Higher Education Funding Council for
England.


\bibliographystyle{mn2e}
\bibliography{sedbib}

\begin{thebibliography}{92}
\expandafter\ifx\csname natexlab\endcsname\relax\def\natexlab#1{#1}\fi

\bibitem[{{Antonucci} \& {Miller}(1985)}]{antonucci85}
{Antonucci} R.~R.~J., {Miller} J.~S., 1985, \apj, 297, 621

\bibitem[{{Avni} \& {Tananbaum}(1986)}]{avni86}
{Avni} Y., {Tananbaum} H., 1986, \apj, 305, 83

\bibitem[{{Becker} {et~al}\mbox{.}(1995){Becker}, {White}, \&
  {Helfand}}]{FIRST}
{Becker} R.~H., {White} R.~L., {Helfand} D.~J., 1995, \apj, 450, 559

\bibitem[{{Belloni}(2010)}]{belloni10}
{Belloni} T.~M., 2010, in Lecture Notes in Physics, Berlin Springer Verlag,
  Vol. 794, Lecture Notes in Physics, Berlin Springer Verlag, {Belloni} T.,
  ed., p.~53

\bibitem[{{Bian} {et~al}\mbox{.}(2005){Bian}, {Yuan}, \& {Zhao}}]{bian05}
{Bian} W., {Yuan} Q., {Zhao} Y., 2005, \mnras, 364, 187

\bibitem[{{Brinkmann} {et~al}\mbox{.}(2000){Brinkmann}, {Laurent-Muehleisen},
  {Voges}, {Siebert}, {Becker}, {Brotherton}, {White}, \&
  {Gregg}}]{brinkmann00}
{Brinkmann} W., {Laurent-Muehleisen} S.~A., {Voges} W., {Siebert} J., {Becker}
  R.~H., {Brotherton} M.~S., {White} R.~L., {Gregg} M.~D., 2000, \aap, 356, 445

\bibitem[{{Calderone} {et~al}\mbox{.}(2012){Calderone}, {Sbarrato}, \&
  {Ghisellini}}]{calderone12}
{Calderone} G., {Sbarrato} T., {Ghisellini} G., 2012, \mnras, 425, L41

\bibitem[{{Cardelli} {et~al}\mbox{.}(1989){Cardelli}, {Clayton}, \&
  {Mathis}}]{cardelli89}
{Cardelli} J.~A., {Clayton} G.~C., {Mathis} J.~S., 1989, \apj, 345, 245

\bibitem[{{Casali} {et~al}\mbox{.}(2007){Casali}, {Adamson}, {Alves de
  Oliveira}, {Almaini}, {Burch}, {Chuter}, {Elliot}, {Folger}, {Foucaud},
  {Hambly}, {Hastie}, {Henry}, {Hirst}, {Irwin}, {Ives}, {Lawrence}, {Laidlaw},
  {Lee}, {Lewis}, {Lunney}, {McLay}, {Montgomery}, {Pickup}, {Read}, {Rees},
  {Robson}, {Sekiguchi}, {Vick}, {Warren}, \& {Woodward}}]{casali07}
{Casali} M. {et~al.}, 2007, \aap, 467, 777

\bibitem[{{Cutri} {et~al}\mbox{.}(2003){Cutri}, {Skrutskie}, {van Dyk},
  {Beichman}, {Carpenter}, {Chester}, {Cambresy}, {Evans}, {Fowler}, {Gizis},
  {Howard}, {Huchra}, {Jarrett}, {Kopan}, {Kirkpatrick}, {Light}, {Marsh},
  {McCallon}, {Schneider}, {Stiening}, {Sykes}, {Weinberg}, {Wheaton},
  {Wheelock}, \& {Zacarias}}]{2MASSPSC}
{Cutri} R.~M. {et~al.}, 2003, {2MASS All Sky Catalog of point sources.},
  {Cutri, R.~M., Skrutskie, M.~F., van Dyk, S., Beichman, C.~A., Carpenter,
  J.~M., Chester, T., Cambresy, L., Evans, T., Fowler, J., Gizis, J., Howard,
  E., Huchra, J., Jarrett, T., Kopan, E.~L., Kirkpatrick, J.~D., Light, R.~M.,
  Marsh, K.~A., McCallon, H., Schneider, S., Stiening, R., Sykes, M., Weinberg,
  M., Wheaton, W.~A., Wheelock, S., \& Zacarias, N.}, ed.

\bibitem[{{Cutri} {et~al}\mbox{.}(2012){Cutri}, {Wright}, \& {et
  al.}}]{wise_all_sky}
{Cutri} R.~M., {Wright} E.~L., {et al.}, 2012, {Explanatory Supplement to the
  WISE All-Sky Data Release Products}

\bibitem[{{Czerny} \& {Elvis}(1987)}]{czerny87}
{Czerny} B., {Elvis} M., 1987, \apj, 321, 305

\bibitem[{{Dickey} \& {Lockman}(1990)}]{dickey90}
{Dickey} J.~M., {Lockman} F.~J., 1990, \araa, 28, 215

\bibitem[{{Done} {et~al}\mbox{.}(2012){Done}, {Davis}, {Jin}, {Blaes}, \&
  {Ward}}]{done12}
{Done} C., {Davis} S.~W., {Jin} C., {Blaes} O., {Ward} M., 2012, \mnras, 420,
  1848

\bibitem[{{Elvis} {et~al}\mbox{.}(2012){Elvis}, {Hao}, {Civano}, {Brusa},
  {Salvato}, {Bongiorno}, {Capak}, {Zamorani}, {Comastri}, {Jahnke}, {Lusso},
  {Mainieri}, {Trump}, {Ho}, {Aussel}, {Cappelluti}, {Cisternas}, {Frayer},
  {Gilli}, {Hasinger}, {Huchra}, {Impey}, {Koekemoer}, {Lanzuisi}, {Le Floc'h},
  {Lilly}, {Liu}, {McCarthy}, {McCracken}, {Merloni}, {Roeser}, {Sanders},
  {Sargent}, {Scoville}, {Schinnerer}, {Schiminovich}, {Silverman},
  {Taniguchi}, {Vignali}, {Urry}, {Zamojski}, \& {Zatloukal}}]{elvis12}
{Elvis} M. {et~al.}, 2012, \apj, 759, 6

\bibitem[{{Elvis} {et~al}\mbox{.}(1994){Elvis}, {Wilkes}, {McDowell}, {Green},
  {Bechtold}, {Willner}, {Oey}, {Polomski}, \& {Cutri}}]{elvis94}
{Elvis} M. {et~al.}, 1994, \apjs, 95, 1

\bibitem[{{Esin} {et~al}\mbox{.}(1997){Esin}, {McClintock}, \&
  {Narayan}}]{esin97}
{Esin} A.~A., {McClintock} J.~E., {Narayan} R., 1997, \apj, 489, 865

\bibitem[{{Green} {et~al}\mbox{.}(2009){Green}, {Aldcroft}, {Richards},
  {Barkhouse}, {Constantin}, {Haggard}, {Karovska}, {Kim}, {Kim}, {Vikhlinin},
  {Anderson}, {Mossman}, {Kashyap}, {Myers}, {Silverman}, {Wilkes}, \&
  {Tananbaum}}]{green09}
{Green} P.~J. {et~al.}, 2009, \apj, 690, 644

\bibitem[{{Grupe} {et~al}\mbox{.}(2010){Grupe}, {Komossa}, {Leighly}, \&
  {Page}}]{grupe10}
{Grupe} D., {Komossa} S., {Leighly} K.~M., {Page} K.~L., 2010, \apjs, 187, 64

\bibitem[{{Haardt} \& {Maraschi}(1993)}]{haardt93}
{Haardt} F., {Maraschi} L., 1993, \apj, 413, 507

\bibitem[{{Hambly} {et~al}\mbox{.}(2008){Hambly}, {Collins}, {Cross}, {Mann},
  {Read}, {Sutorius}, {Bond}, {Bryant}, {Emerson}, {Lawrence}, {Rimoldini},
  {Stewart}, {Williams}, {Adamson}, {Hirst}, {Dye}, \& {Warren}}]{hambly08}
{Hambly} N.~C. {et~al.}, 2008, \mnras, 384, 637

\bibitem[{{Hao} {et~al}\mbox{.}(2010){Hao}, {Elvis}, {Civano}, {Lanzuisi},
  {Brusa}, {Lusso}, {Zamorani}, {Comastri}, {Bongiorno}, {Impey}, {Koekemoer},
  {Le Floc'h}, {Salvato}, {Sanders}, {Trump}, \& {Vignali}}]{hao10}
{Hao} H. {et~al.}, 2010, \apjl, 724, L59

\bibitem[{{Hao} {et~al}\mbox{.}(2012){Hao}, {Elvis}, {Civano}, {Zamorani},
  {Ho}, {Comastri}, {Bongiorno}, {Merloni}, {Brusa}, {Trump}, {Salvato},
  {Impey}, {Koekemoer}, {Lanzuisi}, {Celotti}, {Jahnke}, {Vignali},
  {Silverman}, {Urry}, {Schawinski}, \& {Capak}}]{hao12}
{Hao} H. {et~al.}, 2012, ArXiv e-prints

\bibitem[{{Hewett} {et~al}\mbox{.}(2006){Hewett}, {Warren}, {Leggett}, \&
  {Hodgkin}}]{hewett06}
{Hewett} P.~C., {Warren} S.~J., {Leggett} S.~K., {Hodgkin} S.~T., 2006, \mnras,
  367, 454

\bibitem[{{Hodgkin} {et~al}\mbox{.}(2009){Hodgkin}, {Irwin}, {Hewett}, \&
  {Warren}}]{hodgkin09}
{Hodgkin} S.~T., {Irwin} M.~J., {Hewett} P.~C., {Warren} S.~J., 2009, \mnras,
  394, 675

\bibitem[{{Hopkins} {et~al}\mbox{.}(2007){Hopkins}, {Richards}, \&
  {Hernquist}}]{hopkins07}
{Hopkins} P.~F., {Richards} G.~T., {Hernquist} L., 2007, \apj, 654, 731

\bibitem[{{Jin} {et~al}\mbox{.}(2012{\natexlab{a}}){Jin}, {Ward}, \&
  {Done}}]{jin12c}
{Jin} C., {Ward} M., {Done} C., 2012{\natexlab{a}}, \mnras, 425, 907

\bibitem[{{Jin} {et~al}\mbox{.}(2012{\natexlab{b}}){Jin}, {Ward}, {Done}, \&
  {Gelbord}}]{jin12a}
{Jin} C., {Ward} M., {Done} C., {Gelbord} J., 2012{\natexlab{b}}, \mnras, 420,
  1825

\bibitem[{{Just} {et~al}\mbox{.}(2007){Just}, {Brandt}, {Shemmer}, {Steffen},
  {Schneider}, {Chartas}, \& {Garmire}}]{just07}
{Just} D.~W., {Brandt} W.~N., {Shemmer} O., {Steffen} A.~T., {Schneider} D.~P.,
  {Chartas} G., {Garmire} G.~P., 2007, \apj, 665, 1004

\bibitem[{{Kellermann} {et~al}\mbox{.}(1989){Kellermann}, {Sramek}, {Schmidt},
  {Shaffer}, \& {Green}}]{kellermann89}
{Kellermann} K.~I., {Sramek} R., {Schmidt} M., {Shaffer} D.~B., {Green} R.,
  1989, \aj, 98, 1195

\bibitem[{{Kelly} {et~al}\mbox{.}(2008){Kelly}, {Bechtold}, {Trump},
  {Vestergaard}, \& {Siemiginowska}}]{kelly08}
{Kelly} B.~C., {Bechtold} J., {Trump} J.~R., {Vestergaard} M., {Siemiginowska}
  A., 2008, \apjs, 176, 355

\bibitem[{{Krawczyk} {et~al}\mbox{.}(2013){Krawczyk}, {Richards}, {Mehta},
  {Vogeley}, {Gallagher}, {Leighly}, {Ross}, \& {Schneider}}]{krawczyk13}
{Krawczyk} C.~M., {Richards} G.~T., {Mehta} S.~S., {Vogeley} M.~S., {Gallagher}
  S.~C., {Leighly} K.~M., {Ross} N.~P., {Schneider} D.~P., 2013, \apjs, 206, 4

\bibitem[{{Lawrence} {et~al}\mbox{.}(2007){Lawrence}, {Warren}, {Almaini},
  {Edge}, {Hambly}, {Jameson}, {Lucas}, {Casali}, {Adamson}, {Dye}, {Emerson},
  {Foucaud}, {Hewett}, {Hirst}, {Hodgkin}, {Irwin}, {Lodieu}, {McMahon},
  {Simpson}, {Smail}, {Mortlock}, \& {Folger}}]{lawrence07}
{Lawrence} A. {et~al.}, 2007, \mnras, 379, 1599

\bibitem[{{Lusso} {et~al}\mbox{.}(2012){Lusso}, {Comastri}, {Simmons},
  {Mignoli}, {Zamorani}, {Vignali}, {Brusa}, {Shankar}, {Lutz}, {Trump},
  {Maiolino}, {Gilli}, {Bolzonella}, {Puccetti}, {Salvato}, {Impey}, {Civano},
  {Elvis}, {Mainieri}, {Silverman}, {Koekemoer}, {Bongiorno}, {Merloni},
  {Berta}, {Le Floc'h}, {Magnelli}, {Pozzi}, \& {Riguccini}}]{lusso12}
{Lusso} E. {et~al.}, 2012, \mnras, 425, 623

\bibitem[{{Lusso} {et~al}\mbox{.}(2010){Lusso}, {Comastri}, {Vignali},
  {Zamorani}, {Brusa}, {Gilli}, {Iwasawa}, {Salvato}, {Civano}, {Elvis},
  {Merloni}, {Bongiorno}, {Trump}, {Koekemoer}, {Schinnerer}, {Le Floc'h},
  {Cappelluti}, {Jahnke}, {Sargent}, {Silverman}, {Mainieri}, {Fiore},
  {Bolzonella}, {Le F{\`e}vre}, {Garilli}, {Iovino}, {Kneib}, {Lamareille},
  {Lilly}, {Mignoli}, {Scodeggio}, \& {Vergani}}]{lusso10}
{Lusso} E. {et~al.}, 2010, \aap, 512, A34+

\bibitem[{{Maiolino} {et~al}\mbox{.}(2007){Maiolino}, {Shemmer}, {Imanishi},
  {Netzer}, {Oliva}, {Lutz}, \& {Sturm}}]{maiolino07}
{Maiolino} R., {Shemmer} O., {Imanishi} M., {Netzer} H., {Oliva} E., {Lutz} D.,
  {Sturm} E., 2007, \aap, 468, 979

\bibitem[{{Malkan} \& {Sargent}(1982)}]{malkan82}
{Malkan} M.~A., {Sargent} W.~L.~W., 1982, \apj, 254, 22

\bibitem[{{Marconi} \& {Hunt}(2003)}]{marconi03}
{Marconi} A., {Hunt} L.~K., 2003, \apjl, 589, L21

\bibitem[{{Marconi} {et~al}\mbox{.}(2004){Marconi}, {Risaliti}, {Gilli},
  {Hunt}, {Maiolino}, \& {Salvati}}]{marconi04}
{Marconi} A., {Risaliti} G., {Gilli} R., {Hunt} L.~K., {Maiolino} R., {Salvati}
  M., 2004, \mnras, 351, 169

\bibitem[{{Mateos} {et~al}\mbox{.}(2005{\natexlab{a}}){Mateos}, {Barcons},
  {Carrera}, {Ceballos}, {Caccianiga}, {Lamer}, {Maccacaro}, {Page}, {Schwope},
  \& {Watson}}]{mateos05a}
{Mateos} S. {et~al.}, 2005{\natexlab{a}}, \aap, 433, 855

\bibitem[{{Mateos} {et~al}\mbox{.}(2005{\natexlab{b}}){Mateos}, {Barcons},
  {Carrera}, {Ceballos}, {Hasinger}, {Lehmann}, {Fabian}, \&
  {Streblyanska}}]{mateos05b}
{Mateos} S., {Barcons} X., {Carrera} F.~J., {Ceballos} M.~T., {Hasinger} G.,
  {Lehmann} I., {Fabian} A.~C., {Streblyanska} A., 2005{\natexlab{b}}, \aap,
  444, 79

\bibitem[{{McLeod} \& {Rieke}(1995)}]{mcleod95}
{McLeod} K.~K., {Rieke} G.~H., 1995, \apjl, 454, L77

\bibitem[{{Miller} {et~al}\mbox{.}(2011){Miller}, {Brandt}, {Schneider},
  {Gibson}, {Steffen}, \& {Wu}}]{miller11}
{Miller} B.~P., {Brandt} W.~N., {Schneider} D.~P., {Gibson} R.~R., {Steffen}
  A.~T., {Wu} J., 2011, \apj, 726, 20

\bibitem[{{Morrissey} {et~al}\mbox{.}(2007){Morrissey}, {Conrow}, {Barlow},
  {Small}, {Seibert}, {Wyder}, {Budav{\'a}ri}, {Arnouts}, {Friedman},
  {Forster}, {Martin}, {Neff}, {Schiminovich}, {Bianchi}, {Donas}, {Heckman},
  {Lee}, {Madore}, {Milliard}, {Rich}, {Szalay}, {Welsh}, \& {Yi}}]{galex}
{Morrissey} P. {et~al.}, 2007, \apjs, 173, 682

\bibitem[{{Mullaney} {et~al}\mbox{.}(2012){Mullaney}, {Pannella}, {Daddi},
  {Alexander}, {Elbaz}, {Hickox}, {Bournaud}, {Altieri}, {Aussel}, {Coia},
  {Dannerbauer}, {Dasyra}, {Dickinson}, {Hwang}, {Kartaltepe}, {Leiton},
  {Magdis}, {Magnelli}, {Popesso}, {Valtchanov}, {Bauer}, {Brandt}, {Del Moro},
  {Hanish}, {Ivison}, {Juneau}, {Luo}, {Lutz}, {Sargent}, {Scott}, \&
  {Xue}}]{mullaney12}
{Mullaney} J.~R. {et~al.}, 2012, \mnras, 419, 95

\bibitem[{{Mushotzky} {et~al}\mbox{.}(1980){Mushotzky}, {Marshall}, {Boldt},
  {Holt}, \& {Serlemitsos}}]{mushotzky80}
{Mushotzky} R.~F., {Marshall} F.~E., {Boldt} E.~A., {Holt} S.~S., {Serlemitsos}
  P.~J., 1980, \apj, 235, 377

\bibitem[{{Netzer} {et~al}\mbox{.}(2007){Netzer}, {Lutz}, {Schweitzer},
  {Contursi}, {Sturm}, {Tacconi}, {Veilleux}, {Kim}, {Rupke}, {Baker},
  {Dasyra}, {Mazzarella}, \& {Lord}}]{netzer07}
{Netzer} H. {et~al.}, 2007, \apj, 666, 806

\bibitem[{{Oke} \& {Gunn}(1983)}]{okegunn83}
{Oke} J.~B., {Gunn} J.~E., 1983, \apj, 266, 713

\bibitem[{{Piconcelli} {et~al}\mbox{.}(2003){Piconcelli}, {Cappi}, {Bassani},
  {Di Cocco}, \& {Dadina}}]{piconcelli03}
{Piconcelli} E., {Cappi} M., {Bassani} L., {Di Cocco} G., {Dadina} M., 2003,
  \aap, 412, 689

\bibitem[{{Piconcelli} {et~al}\mbox{.}(2005){Piconcelli}, {Jimenez-Bail{\'o}n},
  {Guainazzi}, {Schartel}, {Rodr{\'{\i}}guez-Pascual}, \&
  {Santos-Lle{\'o}}}]{piconcelli05}
{Piconcelli} E., {Jimenez-Bail{\'o}n} E., {Guainazzi} M., {Schartel} N.,
  {Rodr{\'{\i}}guez-Pascual} P.~M., {Santos-Lle{\'o}} M., 2005, \aap, 432, 15

\bibitem[{{Polletta} {et~al}\mbox{.}(2007){Polletta}, {Tajer}, {Maraschi},
  {Trinchieri}, {Lonsdale}, {Chiappetti}, {Andreon}, {Pierre}, {Le F{\`e}vre},
  {Zamorani}, {Maccagni}, {Garcet}, {Surdej}, {Franceschini}, {Alloin},
  {Shupe}, {Surace}, {Fang}, {Rowan-Robinson}, {Smith}, \&
  {Tresse}}]{polletta07}
{Polletta} M. {et~al.}, 2007, \apj, 663, 81

\bibitem[{{Porquet} {et~al}\mbox{.}(2004){Porquet}, {Reeves}, {O'Brien}, \&
  {Brinkmann}}]{porquet04}
{Porquet} D., {Reeves} J.~N., {O'Brien} P., {Brinkmann} W., 2004, \aap, 422, 85

\bibitem[{{Pounds} {et~al}\mbox{.}(1990){Pounds}, {Nandra}, {Stewart},
  {George}, \& {Fabian}}]{pounds90}
{Pounds} K.~A., {Nandra} K., {Stewart} G.~C., {George} I.~M., {Fabian} A.~C.,
  1990, \nat, 344, 132

\bibitem[{{Rees} {et~al}\mbox{.}(1969){Rees}, {Silk}, {Werner}, \&
  {Wickramasinghe}}]{rees69}
{Rees} M.~J., {Silk} J.~I., {Werner} M.~W., {Wickramasinghe} N.~C., 1969, \nat,
  223, 788

\bibitem[{{Reeves} \& {Turner}(2000)}]{reeves00}
{Reeves} J.~N., {Turner} M.~J.~L., 2000, \mnras, 316, 234

\bibitem[{{Richards} {et~al}\mbox{.}(2003){Richards}, {Hall}, {Vanden Berk},
  {Strauss}, {Schneider}, {Weinstein}, {Reichard}, {York}, {Knapp}, {Fan},
  {Ivezi{\'c}}, {Brinkmann}, {Budav{\'a}ri}, {Csabai}, \&
  {Nichol}}]{richards03}
{Richards} G.~T. {et~al.}, 2003, \aj, 126, 1131

\bibitem[{{Richards} {et~al}\mbox{.}(2006){Richards}, {Lacy},
  {Storrie-Lombardi}, {Hall}, {Gallagher}, {Hines}, {Fan}, {Papovich}, {Vanden
  Berk}, {Trammell}, {Schneider}, {Vestergaard}, {York}, {Jester}, {Anderson},
  {Budav{\'a}ri}, \& {Szalay}}]{richards06}
{Richards} G.~T. {et~al.}, 2006, \apjs, 166, 470

\bibitem[{{Rieke}(1978)}]{rieke78}
{Rieke} G.~H., 1978, \apj, 226, 550

\bibitem[{{Risaliti} {et~al}\mbox{.}(2009){Risaliti}, {Young}, \&
  {Elvis}}]{risaliti09}
{Risaliti} G., {Young} M., {Elvis} M., 2009, \apjl, 700, L6

\bibitem[{{Sanders} {et~al}\mbox{.}(1989){Sanders}, {Phinney}, {Neugebauer},
  {Soifer}, \& {Matthews}}]{sanders89}
{Sanders} D.~B., {Phinney} E.~S., {Neugebauer} G., {Soifer} B.~T., {Matthews}
  K., 1989, \apj, 347, 29

\bibitem[{{Schlegel} {et~al}\mbox{.}(1998){Schlegel}, {Finkbeiner}, \&
  {Davis}}]{schlegel98}
{Schlegel} D.~J., {Finkbeiner} D.~P., {Davis} M., 1998, \apj, 500, 525

\bibitem[{{Schneider} {et~al}\mbox{.}(2007){Schneider}, {Hall}, {Richards},
  {Strauss}, \& {Vanden Berk}}]{DR5QSO}
{Schneider} D.~P., {Hall} P.~B., {Richards} G.~T., {Strauss} M.~A., {Vanden
  Berk} D.~E., 2007, \aj, 134, 102

\bibitem[{{Scott} {et~al}\mbox{.}(2012){Scott}, {Stewart}, \&
  {Mateos}}]{scott12}
{Scott} A.~E., {Stewart} G.~C., {Mateos} S., 2012, \mnras, 423, 2633

\bibitem[{{Scott} {et~al}\mbox{.}(2011){Scott}, {Stewart}, {Mateos},
  {Alexander}, {Hutton}, \& {Ward}}]{scott11}
{Scott} A.~E., {Stewart} G.~C., {Mateos} S., {Alexander} D.~M., {Hutton} S.,
  {Ward} M.~J., 2011, \mnras, 417, 992, (S11)

\bibitem[{{Shang} {et~al}\mbox{.}(2011){Shang}, {Brotherton}, {Wills}, {Wills},
  {Cales}, {Dale}, {Green}, {Runnoe}, {Nemmen}, {Gallagher}, {Ganguly},
  {Hines}, {Kelly}, {Kriss}, {Li}, {Tang}, \& {Xie}}]{shang11}
{Shang} Z. {et~al.}, 2011, \apjs, 196, 2

\bibitem[{{Shemmer} {et~al}\mbox{.}(2008){Shemmer}, {Brandt}, {Netzer},
  {Maiolino}, \& {Kaspi}}]{shemmer08}
{Shemmer} O., {Brandt} W.~N., {Netzer} H., {Maiolino} R., {Kaspi} S., 2008,
  \apj, 682, 81

\bibitem[{{Shemmer} {et~al}\mbox{.}(2006){Shemmer}, {Brandt}, {Schneider},
  {Fan}, {Strauss}, {Diamond-Stanic}, {Richards}, {Anderson}, {Gunn}, \&
  {Brinkmann}}]{shemmer06}
{Shemmer} O. {et~al.}, 2006, \apj, 644, 86

\bibitem[{{Shemmer} {et~al}\mbox{.}(2005){Shemmer}, {Brandt}, {Vignali},
  {Schneider}, {Fan}, {Richards}, \& {Strauss}}]{shemmer05}
{Shemmer} O., {Brandt} W.~N., {Vignali} C., {Schneider} D.~P., {Fan} X.,
  {Richards} G.~T., {Strauss} M.~A., 2005, \apj, 630, 729

\bibitem[{{Shen} {et~al}\mbox{.}(2008){Shen}, {Greene}, {Strauss}, {Richards},
  \& {Schneider}}]{shen08a}
{Shen} Y., {Greene} J.~E., {Strauss} M.~A., {Richards} G.~T., {Schneider}
  D.~P., 2008, \apj, 680, 169

\bibitem[{{Shen} {et~al}\mbox{.}(2011){Shen}, {Richards}, {Strauss}, {Hall},
  {Schneider}, {Snedden}, {Bizyaev}, {Brewington}, {Malanushenko},
  {Malanushenko}, {Oravetz}, {Pan}, \& {Simmons}}]{shen11}
{Shen} Y. {et~al.}, 2011, \apjs, 194, 45

\bibitem[{{Shields}(1978)}]{shields78}
{Shields} G.~A., 1978, \nat, 272, 706

\bibitem[{{Shull} {et~al}\mbox{.}(2012){Shull}, {Stevans}, \&
  {Danforth}}]{shull12}
{Shull} J.~M., {Stevans} M., {Danforth} C.~W., 2012, \apj, 752, 162

\bibitem[{{Spergel} {et~al}\mbox{.}(2003){Spergel}, {Verde}, {Peiris},
  {Komatsu}, {Nolta}, {Bennett}, {Halpern}, {Hinshaw}, {Jarosik}, {Kogut},
  {Limon}, {Meyer}, {Page}, {Tucker}, {Weiland}, {Wollack}, \&
  {Wright}}]{spergel03}
{Spergel} D.~N. {et~al.}, 2003, \apjs, 148, 175

\bibitem[{{Stalin} {et~al}\mbox{.}(2010){Stalin}, {Petitjean}, {Srianand},
  {Fox}, {Coppolani}, \& {Schwope}}]{stalin10}
{Stalin} C.~S., {Petitjean} P., {Srianand} R., {Fox} A.~J., {Coppolani} F.,
  {Schwope} A., 2010, \mnras, 401, 294

\bibitem[{{Steffen} {et~al}\mbox{.}(2006){Steffen}, {Strateva}, {Brandt},
  {Alexander}, {Koekemoer}, {Lehmer}, {Schneider}, \& {Vignali}}]{steffen06}
{Steffen} A.~T., {Strateva} I., {Brandt} W.~N., {Alexander} D.~M., {Koekemoer}
  A.~M., {Lehmer} B.~D., {Schneider} D.~P., {Vignali} C., 2006, \aj, 131, 2826

\bibitem[{{Strateva} {et~al}\mbox{.}(2005){Strateva}, {Brandt}, {Schneider},
  {Vanden Berk}, \& {Vignali}}]{strateva05}
{Strateva} I.~V., {Brandt} W.~N., {Schneider} D.~P., {Vanden Berk} D.~G.,
  {Vignali} C., 2005, \aj, 130, 387

\bibitem[{{Str{\"u}der} {et~al}\mbox{.}(2001){Str{\"u}der}, {Briel}, {Dennerl},
  {Hartmann}, {Kendziorra}, {Meidinger}, {Pfeffermann}, {Reppin}, {Aschenbach},
  {Bornemann}, {Br{\"a}uninger}, {Burkert}, {Elender}, {Freyberg}, {Haberl},
  {Hartner}, {Heuschmann}, {Hippmann}, {Kastelic}, {Kemmer}, {Kettenring},
  {Kink}, {Krause}, {M{\"u}ller}, {Oppitz}, {Pietsch}, {Popp}, {Predehl},
  {Read}, {Stephan}, {St{\"o}tter}, {Tr{\"u}mper}, {Holl}, {Kemmer}, {Soltau},
  {St{\"o}tter}, {Weber}, {Weichert}, {von Zanthier}, {Carathanassis}, {Lutz},
  {Richter}, {Solc}, {B{\"o}ttcher}, {Kuster}, {Staubert}, {Abbey}, {Holland},
  {Turner}, {Balasini}, {Bignami}, {La Palombara}, {Villa}, {Buttler},
  {Gianini}, {Lain{\'e}}, {Lumb}, \& {Dhez}}]{pn}
{Str{\"u}der} L. {et~al.}, 2001, \aap, 365, L18

\bibitem[{{Sutherland} \& {Saunders}(1992)}]{sutherland92}
{Sutherland} W., {Saunders} W., 1992, \mnras, 259, 413

\bibitem[{{Tozzi} {et~al}\mbox{.}(2006){Tozzi}, {Gilli}, {Mainieri}, {Norman},
  {Risaliti}, {Rosati}, {Bergeron}, {Borgani}, {Giacconi}, {Hasinger},
  {Nonino}, {Streblyanska}, {Szokoly}, {Wang}, \& {Zheng}}]{tozzi06}
{Tozzi} P. {et~al.}, 2006, \aap, 451, 457

\bibitem[{{Turner} {et~al}\mbox{.}(2001){Turner}, {Abbey}, {Arnaud},
  {Balasini}, {Barbera}, {Belsole}, {Bennie}, {Bernard}, {Bignami}, {Boer},
  {Briel}, {Butler}, {Cara}, {Chabaud}, {Cole}, {Collura}, {Conte}, {Cros},
  {Denby}, {Dhez}, {Di Coco}, {Dowson}, {Ferrando}, {Ghizzardi}, {Gianotti},
  {Goodall}, {Gretton}, {Griffiths}, {Hainaut}, {Hochedez}, {Holland},
  {Jourdain}, {Kendziorra}, {Lagostina}, {Laine}, {La Palombara}, {Lortholary},
  {Lumb}, {Marty}, {Molendi}, {Pigot}, {Poindron}, {Pounds}, {Reeves},
  {Reppin}, {Rothenflug}, {Salvetat}, {Sauvageot}, {Schmitt}, {Sembay},
  {Short}, {Spragg}, {Stephen}, {Str{\"u}der}, {Tiengo}, {Trifoglio},
  {Tr{\"u}mper}, {Vercellone}, {Vigroux}, {Villa}, {Ward}, {Whitehead}, \&
  {Zonca}}]{MOS}
{Turner} M.~J.~L. {et~al.}, 2001, \aap, 365, L27

\bibitem[{{Vasudevan} \& {Fabian}(2007)}]{vasudevan07}
{Vasudevan} R.~V., {Fabian} A.~C., 2007, \mnras, 381, 1235

\bibitem[{{Vignali} {et~al}\mbox{.}(2003){Vignali}, {Brandt}, \&
  {Schneider}}]{vignali03}
{Vignali} C., {Brandt} W.~N., {Schneider} D.~P., 2003, \aj, 125, 433

\bibitem[{{Vignali} {et~al}\mbox{.}(2005){Vignali}, {Brandt}, {Schneider}, \&
  {Kaspi}}]{vignali05}
{Vignali} C., {Brandt} W.~N., {Schneider} D.~P., {Kaspi} S., 2005, \aj, 129,
  2519

\bibitem[{{Wagner}(2008)}]{wagner08}
{Wagner} R.~M., 2008, \mnras, 385, 119

\bibitem[{{Wang} {et~al}\mbox{.}(2004){Wang}, {Watarai}, \&
  {Mineshige}}]{wang04}
{Wang} J.-M., {Watarai} K.-Y., {Mineshige} S., 2004, \apjl, 607, L107

\bibitem[{{Watson} {et~al}\mbox{.}(2009){Watson}, {Schr{\"o}der}, {Fyfe},
  {Page}, {Lamer}, {Mateos}, {Pye}, {Sakano}, \& {Rosen}}]{2XMM}
{Watson} M.~G. {et~al.}, 2009, \aap, 493, 339

\bibitem[{{Worrall} {et~al}\mbox{.}(1987){Worrall}, {Tananbaum}, {Giommi}, \&
  {Zamorani}}]{worrall87}
{Worrall} D.~M., {Tananbaum} H., {Giommi} P., {Zamorani} G., 1987, \apj, 313,
  596

\bibitem[{{Wyder} {et~al}\mbox{.}(2007){Wyder}, {Martin}, {Schiminovich},
  {Seibert}, {Budav{\'a}ri}, {Treyer}, {Barlow}, {Forster}, {Friedman},
  {Morrissey}, {Neff}, {Small}, {Bianchi}, {Donas}, {Heckman}, {Lee}, {Madore},
  {Milliard}, {Rich}, {Szalay}, {Welsh}, \& {Yi}}]{wyder07}
{Wyder} T.~K. {et~al.}, 2007, \apjs, 173, 293

\bibitem[{{York}(2000)}]{SDSS}
{York} D.~G., 2000, \aj, 120, 1579

\bibitem[{{Young} {et~al}\mbox{.}(2009){Young}, {Elvis}, \&
  {Risaliti}}]{young09}
{Young} M., {Elvis} M., {Risaliti} G., 2009, \apjs, 183, 17

\bibitem[{{Zamorani} {et~al}\mbox{.}(1981){Zamorani}, {Henry}, {Maccacaro},
  {Tananbaum}, {Soltan}, {Avni}, {Liebert}, {Stocke}, {Strittmatter},
  {Weymann}, {Smith}, \& {Condon}}]{zamorani81}
{Zamorani} G. {et~al.}, 1981, \apj, 245, 357

\bibitem[{{Zhou} {et~al}\mbox{.}(2006){Zhou}, {Wang}, {Yuan}, {Lu}, {Dong},
  {Wang}, \& {Lu}}]{zhou06}
{Zhou} H., {Wang} T., {Yuan} W., {Lu} H., {Dong} X., {Wang} J., {Lu} Y., 2006,
  \apjs, 166, 128

\end{thebibliography}


\bsp

\label{lastpage}

\end{document}